\documentclass[a4paper,11pt]{article}
\pdfoutput=1 
\usepackage{jcappub} 
\usepackage[T1]{fontenc} 
\usepackage{graphicx}   
\usepackage{changepage}
\usepackage{academicons}
\usepackage{natbib}
\usepackage{multirow}
\usepackage{orcidlink}
\usepackage{subcaption}
\usepackage{academicons}
\definecolor{orcidlogocol}{HTML}{A6CE39}

\usepackage{aas_macros}
\title{\boldmath{SpectrAx: Spectral Search of Axion-Like Particles Using Multi-Band Observations of Galaxy Clusters  from SKA, SO, CMB-S4 and eROSITA}}

\author{Harsh Mehta\orcidlink{0009-0007-4664-4820},}
\author{and Suvodip Mukherjee\orcidlink{0000-0002-3373-5236}}

\affiliation{Department of Astronomy and Astrophysics, Tata Institute of Fundamental Research, Homi Bhabha Road, Mumbai- 400005, India}

\emailAdd{harsh.mehta@tifr.res.in}
\emailAdd{suvodip.mukherjee@tifr.res.in}

\begin{document}
\abstract{
The existence of axions or Axion-Like Particles (ALPs) has been predicted by various Beyond Standard Model (BSM) theories, and the proposed photon-ALP interaction is one of the ways to probe them. Such an interaction will lead to photon-ALP resonant conversion in galaxy clusters, resulting in a polarized spectral distortion in the CMB along the cluster line of sight. The estimation of this signal from galaxy clusters requires an estimation of their electron density and magnetic field profiles, as well as their redshifts. We have developed a new Bayesian framework \texttt{SpectrAx} that can use observations from different electromagnetic bands such as radio, CMB, optical, and X-ray to infer the astrophysical properties of a galaxy cluster, such as cluster its redshift, electron density and magnetic field, along with the BSM physics such as ALPs. 
We use simulated redshifts in our analysis, but that can be obtained by cross-matching with optical surveys having overlapping sky regions with the galaxy clusters. Also, we use radial profiles that are motivated from observations of galaxy clusters at low redshifts. By using the simulated data corresponding to the ALP mass of $10^{-14}$ eV for upcoming CMB surveys such as Simons Observatory (SO) and CMB-S4 in combination with Square Kilometer Array (SKA) and extended ROentgen Survey with an Imaging Telescope Array (eROSITA) we demonstrate the capability in accurately inferring the ALPs coupling strength along with the radial profile of electron density and magnetic field from galaxy clusters. The application of this framework to the data from future surveys by combining SKA+SO+eROSITA and SKA+CMB-S4+eROSITA will make it possible for the first time to explore both astrophysics and BSM physics from low-redshift galaxy clusters using a multi-band approach.}

\maketitle
\flushbottom

\section{Introduction}
The $\Lambda$CDM universe consists of various components in radiation, matter, and dark energy, with all of these affecting the evolution of the universe during different stages \cite{alpher1948evolution,ratra2008beginning,gamow1948evolution}. The matter component is composed mostly of dark matter ($\sim 85\%$), followed by baryons ($\sim 15\%$), with the rest being non-relativistic neutrinos. 
Dark matter halos are the largest gravitationally bound structures in the universe. These halos are hosts to galaxy clusters, which are the largest observable gravitationally bound structures. Thus, it is these clusters that can help us connect astrophysics to cosmology. These clusters can be observed due to the multiple phenomena giving rise to different kinds of emissions, corresponding to various baryon and baryon-photon interactions \cite{Dodelson:2003ft,brunetti2014cosmic,bonafede2009revealing,eckert2016xxl}.

The clusters are hosts to galaxies, but most of their mass comes from the intracluster medium (ICM) which mainly consists of plasma of ionized gases and electrons at a high temperature of around $10^7$ K. These ionized gases and electrons interact with each other and with photons in different processes such as Bremsstrahlung, Compton scattering, and synchrotron emission, leading to emission or absorption of power at different wavelengths \cite{1986rpa..book.....R}.
 The different phenomena leading to these emissions can be probed by studying the spectral and spatial variation of the emissions \cite{prestage1988cluster,kaastra2004spatially}. The strength of the emissions varies from cluster to cluster and primarily depends on their electron density, mass density, magnetic field, and temperature. Thus, measurements of these signals are indicators of different astrophysical characteristics of galaxy clusters in the Universe. 

The relativistic electrons in the presence of a transverse magnetic field, lose their energies in the form of synchrotron radiation. This radiation is polarized perpendicular to the line of sight and the magnetic field at the radiating location \cite{1986rpa..book.....R}, thus its polarization and intensity can be used to infer the direction as well as the magnitude of the transverse magnetic field. The emission peaks at radio frequencies, which can be used to probe the magnetic field in clusters \cite{carilli2002cluster,GOVONI_2004}.

Due to the presence of hot gas in galaxy clusters, the thermal radiation peaks at X-ray wavelengths. Processes like bremsstrahlung (or line-line emission) and inverse Compton effect lead to X-ray emissions from galaxy clusters \cite{1986rpa..book.....R}. The inverse Compton effect, also called the Sunyaev Zeldovich (SZ) effect in galaxy clusters leads to an absorption of photons at high wavelengths and emission at lower ones \cite{1972CoASP...4..173S}. These processes depend heavily on the electron density and the temperature in the galaxy cluster. Hence, the emission at X-ray frequencies can be used to probe the electron density and temperature of a galaxy cluster \cite{sarazin1986x,Birkinshaw_1999}.

All these processes, combined with thermal radiation from galaxies contribute to the emission at optical and infrared wavelengths. This part of the spectrum is used in galaxy surveys \cite{2014ApJS..210....9B,zehavi2011galaxy}. The emission from galaxies and the intervening ICM can tell us about the redshifts, velocity dispersion, shapes, etc., of clusters \cite{castagne2012deep}. The redshifts can be obtained either using photometric (based on flux in various bands) \cite{abdalla2011comparison} or spectroscopic means (using a study of transition lines) \cite{zaznobin2021spectroscopic}.

The radiation from these clusters forms just a part of the total radiation content of the universe, which is mainly contributed to by the Cosmic Microwave Background (CMB) \cite{Dodelson:2003ft}. The CMB peaks at microwave frequencies and follows an almost ideal black-body spectrum \cite{Fixsen_1996,fixsen1996cosmic}. Also, it is highly isotropic and exhibits anisotropy only 1 part in $10^{5}$ in temperature \cite{Dodelson:2003ft,Fixsen_1996,fixsen1996cosmic,hanson2009estimators}. It is a relic of the Big Bang after the photons decoupled from the photon-baryon fluid at the surface of last scattering. The CMB photons as they travelled through the expanding universe, got cosmologically redshifted to a temperature of 2.7255 K, as is observed today \cite{Fixsen_2009}. 
The anisotropies in the CMB are a result of phenomena in the early universe before recombination (also called primary anisotropies) or after recombination (called secondary anisotropies) \cite{Dodelson:2003ft,hu2002cosmic}. The secondary anisotropies are generated due to the gravitational or electromagnetic (EM) interactions of CMB photons with the intervening matter field in the universe. The CMB photons as they travel through or near a galaxy cluster may get gravitationally lensed due to the cluster mass or may electromagnetically interact with other photons and baryons. This will give rise to secondary anisotropies in the CMB such as reionization \cite{adam2016planck,}, lensing \cite{Smith_2007}, SZ effects \cite{1972CoASP...4..173S}, etc., which can be probed by resolving the temperature and/or polarization fluctuations in the CMB along the cluster line of sight. Also, processes such as the SZ effects will distort the Planck black-body spectrum of the CMB, leading to spectral distortions \cite{2014PTEP.2014fB107T,erler2018planck}.

One such anisotropic spectral distortion in the CMB is caused due to production of axion-like particles (ALPs) \cite{dine1983not,abbott1983cosmological,preskill1983cosmology,Ghosh:2022rta,1992SvJNP..55.1063B,khlopov1999nonlinear,sakharov1994nonhomogeneity,sakharov1996large} around galaxy clusters, if ALPs exist \cite{Mukherjee_2020, Mehta:2024wfo, Mehta:2024pdz}. If axions or axion-like particles (ALPs) exist, the weak coupling they may be having with photons usually denoted by the coupling constant $g_{a\gamma}$ \cite{Carosi:2013rla,Mukherjee_2020,Ghosh:2023xhs}  will lead to the conversion of CMB photons to ALPs as they travel through the cluster.  When the effective masses of the photon and the ALP are the same, the conversion is resonant and leads to a polarized spectral distortion in the CMB along the cluster line of sight. The probability of such a conversion depends on the 
transverse magnetic field, electron density at the resonant location, the redshift of the cluster and the coupling constant, $g_{a\gamma}$ \cite{Mukherjee_2018,Mukherjee_2019,Mukherjee_2020,osti_22525054}. 

In this work, we have developed a Bayesian data analysis framework that can use observations from different bands such as radio, CMB, optical, and X-rays to infer the cluster magnetic field, electron density, and redshift from observations and combine those measurements (and uncertainties) to obtain the coupling strength between ALPs and photon, $g_{a\gamma}$. The various galaxy cluster locations can be identified using observations in optical wavelengths. 
The redshift measurements from these bands will be included and cross-matched with data from other frequency surveys \cite{salvato2018finding}.
The magnetic field in galaxy clusters can be constrained using synchrotron emission from radio observations. This gives us the transverse magnetic field along the line of sight. The X-ray observations can be used to estimate the electron density and temperature profiles in the cluster. All these measurements will be incorporated into the hierarchical Bayesian framework (see Section \ref{sec:bayes_axion}) \texttt{\texttt{SpectrAx}}, which can be applied on the foreground cleaned CMB maps to obtain constraints on the ALPs coupling constant for various masses. 

The motivation behind developing the multi-band framework is highlighted in Sec.  \ref{sec:motive}. This is followed by a review of the various emissions and the photon-ALP resonant conversion phenomenon relevant for \texttt{SpectrAx} in Sec.  \ref{sec:emissions}. The components of \texttt{SpectrAx} and their working is explained in Sec.  \ref{sec:method}. The results on the coupling constant, obtained from mock data using \texttt{SpectrAx} are showcased in Sec.  \ref{sec:results}.
The multi-band framework and its significance in the upcoming era are summarized in Sec.  \ref{sec:conclusion}. We have mostly used natural units ($\hbar = 1, c = 1, k_B = 1$), until explicitly mentioned and the cosmological parameters from Planck 2015 results \cite{2016} (Planck TT, TE, EE + low P + Lensing data) for this analysis.

\section{Motivation}\label{sec:motive}

\begin{figure}[h!]
     \centering
\includegraphics[height=4.5cm,width=17cm]{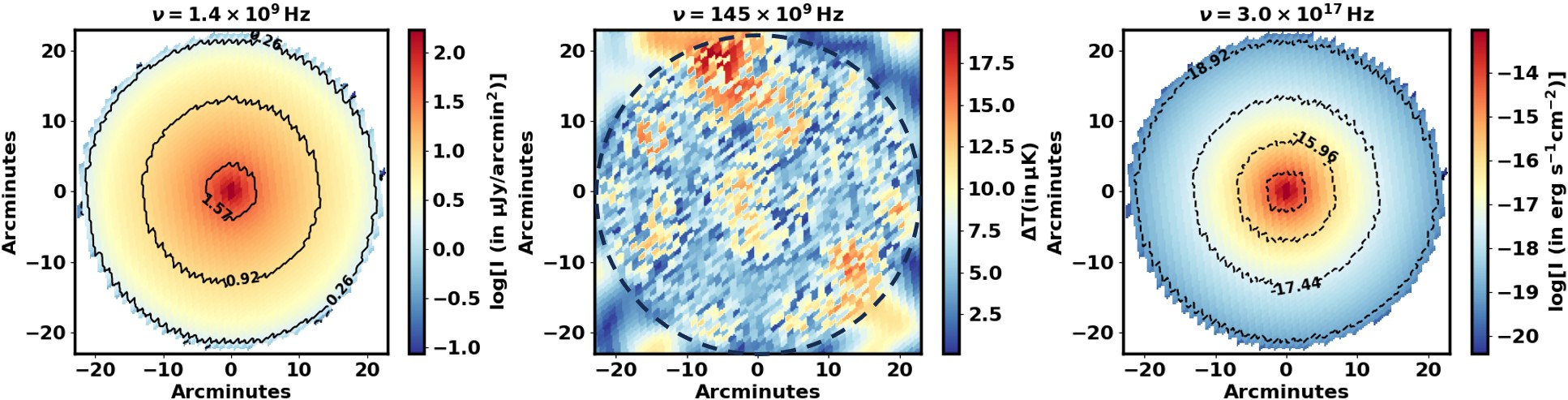}
\caption{The emission from a cluster at different frequencies. The cluster emits mainly through synchrotron radiation on the left at radio frequencies, and through thermal X-ray emission on the right at high frequencies. The emission at the center shows the cluster at the microwave frequency of 145 GHz, if ALPs exist in nature. This will lead to an ALP distortion in the CMB in the cluster region. The black dashed curve encloses the resolved cluster line of sight. The scale is same for all the three subfigures.} 
\label{fig: bands}
\end{figure}

The galaxy clusters are the largest observable gravitationally bound structures in the universe, and act as a bridge between astrophysics and cosmology. The intracluster medium (ICM) and the galaxies are sites for many electromagnetic phenomena occurring in clusters \cite{mohr1999properties}. These lead to emissions that cover a broad range of the EM spectrum. Fig. \ref{fig: bands} shows the emission from a galaxy cluster at various frequencies (low to high). The cluster emits mainly through synchrotron radiation on the left at radio frequencies, and through thermal X-ray emission on the right at high frequencies. The emission at the center shows the cluster at the microwave frequency of 145 GHz, if ALPs exist in nature. This will lead to an ALP distortion in the CMB in the cluster region. The angular scale is the same for all the three frequencies.

\begin{figure}[h!]
     \centering
\includegraphics[height=14cm,width=14.5cm]{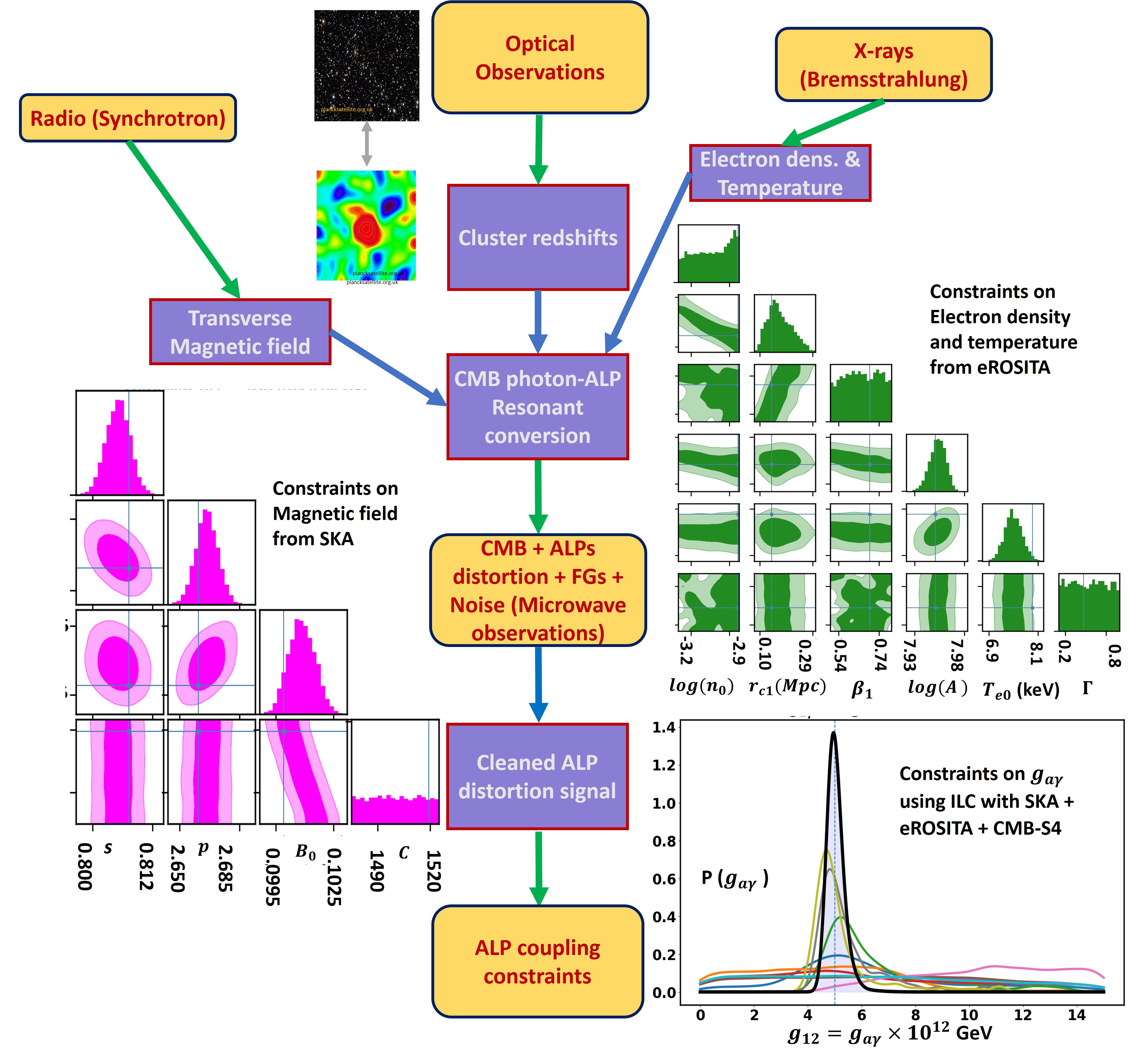}
\caption{This flowchart depicts the working of \texttt{SpectrAx}. \texttt{SpectrAx} infers cluster transverse magnetic fields from radio observations and electron density and temperature profiles from X-rays. These profiles and cluster redshifts determine the ALP distortion signal that is generated in galaxy clusters. This signal is probed from CMB observations of cluster regions, that are contaminated with foregrounds and instrument noise. \texttt{SpectrAx} can clean this ALP signal and using redshifts from cross-matching with optical surveys, \texttt{SpectrAx} is able to obtain constraints on the strength of the ALP coupling with photons.}
\label{fig: flowchart}
\end{figure}

The framework \texttt{SpectrAx}, developed in this work brings the astrophysics of clusters and the cosmological search for axions or ALPs (one of the dark matter candidates), under one umbrella, using multi-band observations of galaxy clusters. Since various emissions from a cluster depend on different properties of a cluster, all the cluster attributes cannot be inferred from a single frequency band, dominated by just a particular kind of emission. Multi-frequency analysis enables a study of different emissions from clusters. The ALP signal generated in galaxy clusters depends on the magnetic field and electron density profiles, and the redshift of the cluster, which can be inferred across different frequency surveys (see Table \ref{tab:prop}). 
The optical observations can provide the redshift information, while the thermal X-ray emission can probe the electron density and temperature profiles of a cluster. The radio emissions from synchrotron will provide constraints on the transverse magnetic field profile. Using CMB observations (contaminated with foregrounds and instrument noise) of cluster regions, and cleaning them, the ALP signal can be probed to obtain constraints on the photon-ALP coupling constant.
 The inferred properties from different bands can then be combined under the hierarchical Bayesian framework to get constraints on the coupling constants of ALPs (see Fig. \ref{fig: flowchart}). 

\begin{table}[h!]
    \centering
{\begin{tabular}{|l|l|l|l|}
    \hline
    \textit{\textbf{ Probes}} & \textit{\textbf{Bandwidth}} & \textit{\textbf{Properties ascertained}} & \textit{\textbf{Considered in \texttt{SpectrAx}} } 
    \tabularnewline \hline
    Radio & 0.7 to 15 GHz & Magnetic field &  \begin{tabular}{@{}c@{}}Smooth radial profiles, \\ Synchrotron intensity\end{tabular}\\
    \hline
    CMB & 27 to 280 GHz &  ALP distortion &
     \begin{tabular}{@{}c@{}}Smooth radial profiles, \\ Foregrounds,\\
     Variable ALP mass \& coupling \end{tabular}\\
    \hline
    Optical & 400 to 700 THz &  
     Cluster redshift & Simulated Redshifts\\
    \hline
    X-ray & 0.5 to 10 keV & Electron density, Temperature & \begin{tabular}{@{}c@{}} Smooth radial profiles, \\ Thermal Bremsstrahlung\end{tabular}\\
    \hline
    \end{tabular} }
    \caption{Astrophysical properties inferred by \texttt{SpectrAx} from multi-probe observations.}
    \label{tab:prop}
    \end{table}

\texttt{SpectrAx} aims to use the data from a number of different types of detectors.  
The upcoming missions like the Simons Observatory (SO) \cite{Ade_2019} and CMB-S4 \cite{abazajian2016cmbs4} will further enrich us with information about the CMB distortions and anisotropies as they will be able to detect fluctuations of the order of $10^{-8}$ K and we intend to use \texttt{SpectrAx} to probe new physics using these detectors. 
With improvement in sensitivity and resolution, upcoming detectors like the Square Kilometer Array (SKA) \cite{Carilli_2004,braun2019anticipated}, Atacama Large Millimeter/submillimeter Array (ALMA) \cite{brown2004alma}, and eROSITA \cite{predehl2021erosita} will be able to observe the universe at different wavelengths. 

There will be a large volume of observational data from multiple bands (like from radio observations by SKA, from X-ray frequencies by eROSITA, from optical and infrared wavelengths by James Webb Space Telescope  (JWST) \cite{gardner2006james}, etc.), which \texttt{SpectrAx} can use to increase our astrophysical understanding of the properties of clusters. This makes it possible to use complementary information from different surveys and search for other secondary anisotropies (like lensing and SZ effect) as well, which are caused due to various phenomena in clusters. As a result, this pipeline can be used for a broad range of science goals including astrophysics and cosmology by combining multi-band observations from SKA, eROSITA, SO, and CMB-S4, and in the future other high resolution CMB experiments such as CMB-HD \cite{sehgal2019cmbhd}.

\section{Emissions from Galaxy clusters}\label{sec:emissions}
\subsection{Magnetic field from radio observations}
\label{sec:radio}
The magnetic field generation in galaxy clusters is possibly from gravitational energy being injected into the ICM. The amplification of the magnetic field takes place via cluster mergers. These magnetic fields affect the cluster from its stellar rotations to plasma hydrodynamics \cite{GOVONI_2004,condon2016essential,carilli2002cluster,clarke2001new,Ferrari_2008}. Radio observations give us information about the magnetic field in a cluster and the accompanying processes such as synchrotron emission and Faraday rotation. The magnetic fields in galaxy clusters are of the order of $\mathrm{ \mu G}$ estimated from various observations. The radiation obtained may or may not be polarized depending on the intrinsic source depolarization and the dominant depolarization by the intracluster medium (ICM) \cite{GOVONI_2004,condon2016essential}. 

The line of sight magnetic field ($B_{||}$) values are obtained using Faraday rotation measurements. The direction of the magnetic field in a medium determines the preferred direction of gyration for the electrons. Since a linear polarization can be decomposed into left and right circular polarizations, this causes a phase shift between the two polarizations. The angle of linear polarization gets rotated due to different refractive indices for the left and right circular polarizations in the presence of a longitudinal magnetic field, that causes a phase shift \cite{ghatak2009optics}. This rotates the plane of polarization as: $\Delta \chi = RM \lambda^2$, with $\lambda$ being the wavelength of photon and the Rotation Measurement ($RM$) is given as \cite{GOVONI_2004,murgia2004magnetic,bohringer2016cosmic,Ferrari_2008,clarke2001new,eilek2002magnetic,bonafede2010galaxy}:
\begin{equation}
RM = 811.9\int_{0}^{L} n_eB_{||}dl \, \mathrm{rad/m^2}.   
\label{eq:FR}
\end{equation}

 The radiation from radio halos is unpolarized, especially due to the depolarization from the dense central regions of the ICM with high electron densities.  The radiation from radio relics exhibits polarization as they are in the outer regions of the cluster. Turbulence and randomness in the magnetic field further lead to depolarization. The cooling flow clusters exhibit higher values of RM measurements and magnetic fields, as the fields are frozen in a higher gas density than that in merging systems \cite{GOVONI_2004,carilli2002cluster,Ferrari_2008}. 

Apart from the Faraday rotation, synchrotron emission from galaxy clusters happens due to the magnetic fields. The synchrotron is the emission by relativistic particles as they move around the transverse magnetic field. The magnetic field and the electron energy determine the critical frequency of the emission, which is the frequency at which the emission is maximum. The clusters exhibit diffused synchrotron emission which can be studied alongside the inverse Compton emission from X-rays to obtain magnetic field strengths.  The magnetic field from the synchrotron emission is obtained by considering the equipartition of energy between particles and that due to the magnetic field.
The energy in a synchrotron halo can be partitioned between those of particle electrons and heavier particles, and that due to the magnetic field. When their sum ($U$) is minimized, we obtain the equipartition of energies and the $B_{\mathrm{eq}}$ is given as \cite{GOVONI_2004,bohringer2016cosmic,clarke2001new,Ferrari_2008}:
\begin{equation}
B_{\mathrm{eq}} = (24\pi U_{\mathrm{min}}/7)^{1/2} .
\label{eq:equi}
\end{equation}

The synchrotron emits radiation within a half-angle $\gamma^{-1}$ about the electron velocity at a point, where $\gamma = (1 - v^2/c^2)^{-1/2}$. The total intensity spectrum for homogeneous and isotropic distribution of electrons ($N(\epsilon)= N_0\epsilon^{-p}$, with $\epsilon$ being the electron energies) is given as: $S(\nu) \propto \nu^{(1 - p)/2}$.
For low frequencies, when the medium becomes optically thick, self-absorption occurs and the intensity varies as,
$\mathrm{S(\nu) \propto \nu^{5/2}}$ \cite{1986rpa..book.....R,condon2016essential,Ferrari_2008}, although this variation is difficult to probe from observations.

There is diffused synchrotron emission of radio frequencies in galaxy clusters. These contain information about the transverse magnetic field along the line of sight. Hence, the transverse magnetic field profiles of the clusters can be estimated using synchrotron emission.  
The total power per unit volume per unit frequency for the power law distribution of electrons follows  \cite{1986rpa..book.....R,condon2016essential}:
\begin{equation}
P_{\mathrm{tot}}(\nu) =  CB^{(1 + p)/2} \nu ^{(1 - p)/2} ,   
\label{eq: pow sync}
\end{equation}
where $C$ is the proportionality factor that depends on the energy distribution. 
The intensity of radio emission from galaxy clusters is about $\sim 1 \, \mu \mathrm{Jy / arcsec^2}$ \cite{loi2017magnetic}.

For our analysis, we consider a radial profile for the magnetic field in galaxy clusters that decays in the outer regions of the cluster. The magnetic field profile considered is motivated from observations of low redshift galaxy clusters (see Equ. \eqref{eq:mag prof}), but SpectrAx will be able to employ complicated magnetic field models as well in its code and estimate the parameters. The addition of nuisance parameters to the profiles can take into account the uncertainty in their modeling. This would make the analysis robust, but with increased computational cost. The addition of turbulence and incoherence would further add to this uncertainty, but these effects will be considered in a future analysis.  

 We have used different parameter values for different clusters within the allowed range for galaxy clusters. 
We calculate the line of sight integrated synchrotron intensity (using Equ.\eqref{eq: pow sync}) at scales corresponding to the pixel resolution and smooth with the beam size of the instrument. Depolarization occurs if the beam size is greater than the magnetic field coherence scale. We assume that is not the case with upcoming high resolution experiments like SKA. 
This is done at different frequency bands to obtain a multi-frequency-based magnetic field inference. The instrument noise is added corresponding to the frequency bands to generate a radio synchrotron emission map for galaxy clusters. The intensity values in the cluster region will provide us with the data vector for our Bayesian analysis. This will allow us to perform a radial tomographic reconstruction of the magnetic field profile using radio observations. This analysis will enable a better understanding of magnetic fields in galaxy clusters and their coherence scales. This would be key in being able to estimate the 3-dimensional profile for non-spherical clusters. The inference of a three-dimensional magnetic field taking into account the coherence, as well as the turbulence in the magnetic field using synchrotron, as well as the Faraday rotation measurements will be included in the \texttt{SpectrAx} framework in the future. In this analysis, we focus on the inference of radial transverse magnetic field using synchrotron emission, while the effects of turbulence and incoherence will be considered in a future work.

\subsection{Electron density and Temperature from X-rays}
\label{sec:xray}
Using X-rays, the clusters are identified as extended sources consisting of the intracluster medium (ICM) \cite{bulbul2024srgerosita,merloni2012erosita,pratt2016hot,mcdonald2013growth,mcdonald2017remarkable}. This is because of the thermal emission from ionized plasma at high temperatures. 
For a gas in hydrostatic equilibrium in the potential well of galaxy cluster, the electron temperature $T_e$ is given as \cite{felten1966x,sarazin1986x,cavaliere1978distribution}:
\begin{equation}
  k_B T_e \approx GM m_p/2R_{\mathrm{eff}},  
\label{eq:elec temp}
\end{equation}
here $R_\mathrm{{eff}}$ and $M$ are the effective radius and mass of the cluster respectively and $m_p$ is the mass of the hydrogen atom. The temperature $T_e$ for galaxy clusters is of the order of tens of millions of kelvins. 
{
The X-ray emission owes its origin to the acceleration of electrons to high energies at high temperatures, i.e., thermal bremsstrahlung. The electrons are accelerated by the electric field of the ions, with the velocities following a Maxwellian velocity distribution. The spectrum follows an exponential decay with frequency, due to the limiting velocity cut-off for photon generation.} 
For the mass and radius of a cluster, the thermal emission is mostly as X-ray, produced through thermal bremsstrahlung (dominant) and line radiation \cite{1986rpa..book.....R}. This leads to hard X-ray or even gamma photons being emitted from clusters. 
The X-ray emission can be written as \cite{sarazin1986x} 
\begin{equation}
I_X = \int \int A n_e^2 T_e^{-1/2} \exp(-h\nu / k_B T_e)\mathrm{d}l,
\label{eq:xray}\end{equation}
where the factor $A$ takes into account the Gaunt factor (that depends on the energy distribution, scattering velocities, and impact parameters), redshift, etc. of the cluster \cite{1986rpa..book.....R}. The X-ray emission from clusters is of the order of $10^{-13}$ to $10^{-11} \, \mathrm{erg \, cm^{-2} \, s^{-1}}$ \cite{cavaliere1978distribution,Vikhlinin_2005,Vikhlinin_2006}. The photon counts in the various bands are used to calculate the X-ray emission. The bands are characterized as soft ($< 2$ keV) and hard ($> 2$ keV) X-ray bands. Galaxy clusters, supernovae, etc. contribute mainly to the soft band, while Active Galactic Nuclei (AGNs) and X-ray binaries are sources of hard X-rays \cite{fadley2014some}. 

For our analysis, we consider radial profiles for the electron density and temperature field in galaxy clusters that decay in the outer regions of the cluster. We calculate the line of sight integrated X-ray intensity (using Equ.\eqref{eq:xray}) at scales corresponding to the pixel resolution and smooth with the beam size of the instrument. This is done at different frequency bands to obtain a multi-frequency-based inference. The instrument noise is added corresponding to the frequency bands to generate a thermal X-ray emission map for galaxy clusters. The intensity values in the cluster region will provide us with the data vector for a Bayesian analysis. This will let us perform a radial tomographic reconstruction of the electron density and temperature profiles from X-rays.  

In addition to X-rays, information on the electron densities and temperature profiles can be obtained using the SZ-effect in clusters \cite{Birkinshaw_1999,Staniszewski_2009,komatsu1999submillimeter}. 
 The inverse Compton effect refers to the scattering of low-energy photons by high-energy electrons, to high energies. 
The frequency change is given as \cite{1986rpa..book.....R}
\begin{equation}
  \Delta \nu =  \frac{4}{3}\gamma^2 \nu_{\mathrm{in}}  ,
\label{eq:changefreq}
\end{equation}
where $\nu_{\mathrm{in}}$ is the incident photon frequency and $\gamma = (1 - v^2 / c^2)^{-1/2}$, with $v$ being the velocity of electrons.
The Comptonization y-parameter for the thermal SZ effect is given as \cite{sunyaev1980microwave}:
\begin{equation}
y = \int n_e \sigma_T (k_B T_e/m_ec^2) dl.
\label{eq: y parameter}
\end{equation}
For low energy photons, the Thomson scattering takes place with optical depth $\tau \sim n_e \sigma_T R_{\mathrm{eff}} \sim 10^{-2}$ to $10^{-3}$ \cite{battaglia2016tau}. The SZ is a redshift-independent effect and can be used to identify clusters even at very high redshifts as it causes a fractional change in CMB brightness. The SZ effect leads to an increase in brightness at higher frequencies, while a decrease at lower frequencies. The spectral dependence of the temperature fluctuations from SZ effect is given as (where $x \equiv h\nu / k_B T_{\mathrm{cmb}}$ ) \cite{sunyaev1980microwave,carlstrom2002cosmology,Komatsu_1999,1972CoASP...4..173S}:
\begin{equation}
g(x) = x\coth(x/2) - 4.   
\end{equation}
The varying dependence of X-rays and SZ effect on the electron density and temperature profiles can break the degeneracy in the inferred electron density and temperature amplitudes \cite{Birkinshaw_1999,adam2017mapping,shitanishi2018thermodynamic,komatsu1999submillimeter}. Mass measurements from lensing can also improve the constraints on these profiles. We have not considered the SZ and mass measurements in the current analysis, and can be included in \texttt{SpectaAx} framework in the future.

\subsection{Cross-matching with the optical survey}\label{sec:optical}
For the multi-band functionality, we need to do a cross-matching of the data from various surveys. The cross-matching involves the matching of declination and right ascension values of a cluster on various frequency sky maps \cite{bhatiani2022optical}. This is required so that information on the cluster from various frequency bands can be obtained. The 3-dimensional localization of a cluster is not possible until its redshift information is available.

The redshifts are generally obtained from optical surveys. There are spectroscopic (based on transition lines in the spectrum) and photometric (based on flux in various bands) methods of redshift estimation \cite{csabai2003application}. The photo-z redshift estimations involve uncertainties with the error on redshift being proportional to $1 + z$ \cite{bilicki2016wise,1972CoASP...4..173S}. The spectroscopic errors are low, but these surveys are limited by specific intensities from the cluster. 

The clusters are identified as overdensities on the colour-magnitude diagrams.
Generally a colour sequence method (mainly red sequence )\cite{bilicki2016wise,gilbank2004exploring,yee2001optical} is used for redshift estimation that identifies a cluster by the
presence of old red galaxies (following a burst of star-formation and passive
evolution) that follow a certain colour-magnitude relation. The identification of
the brightest cluster galaxy (BCG) can also be used to identify clusters as being
overdensities of galaxies around the BCG \cite{wen2018catalogue}. If multiple galaxies from a cluster are
selected as BCG candidates, then conditions on redshift difference and projected
distance can be imposed. 

The morphology and ellipticity of clusters can also be inferred from optical observations.
Optical surveys are also used to provide us with velocity dispersion measurements which gives us constraints on the masses of clusters.
The mass of a cluster can also be found using observations from optical, X-rays, and lensing \cite{girardi1998optical,Vikhlinin_2006,Smith_2007}. The mass cross-matching will provide constraints on the mass of a cluster, but is currently not considered in the analysis. This can be included in the next version of this pipeline.

\subsection{Axion-Like Particles} \label{sec:axions}

Axions or ALPs (axion-like particles) are hypothetical particles, predicted by various beyond Standard model theories. Axions belong to the class of pseudo Nambu-Goldstone bosons (PNGBs) which possess mass. These were first proposed to be produced as solutions to the strong CP problem \cite{berezhiani1991cosmology}.  The global Peccei-Quinn chiral U(1) symmetry breaking (which is not exact at low energies) by the vacuum expectation value of a scalar field leads to the production of axions which possess a small mass, which also makes them one of the dark matter candidates \cite{dine1983not,abbott1983cosmological,preskill1983cosmology,Ghosh:2022rta,1992SvJNP..55.1063B,khlopov1999nonlinear,sakharov1994nonhomogeneity,sakharov1996large,chadha2022axion}.

Similar to the case of neutrino oscillations, axion-photon oscillations are possible if axions can weakly interact with photons \cite{Raffelt:1996wa, Choi_2021,smirnov2005msw,langacker1987implications,forero2014neutrino,duan2010collective,wolfenstein2018neutrino,kuo1989neutrino,bilenky1987massive,bilenky1999phenomenology}. Since axions require high energies for production in particle-based experiments, detecting them using the weak coupling they may be having with photons makes for a suitable alternative  \cite{Carosi:2013rla,Mukherjee_2020,Ghosh:2023xhs,marsh2016axion}. Galaxy clusters provide a way of probing ALP masses of a certain range ($\sim 10^{-15} - 10^{-11}$ eV), but there is a wide range of ALP masses that can be probed using other astrophysical systems as well, such as neutrino halos, galactic halos, etc., \cite{perna2012signatures,bondarenko2023neutron,lella2023protoneutron,yu2023searching}.

The strength of the interaction between a photon and ALP is given by a photon-ALP coupling ($g_{a\gamma}$), the variation of which with axion mass depends on the model. But in the case of ALPs, this variation is not well constrained. The bounds from COBE FIRAS on the product of magnetic field and coupling constant are at $\mathrm{ g_{a\gamma} B < 10^{-13} \, GeV^{-1} \, nG }$ \cite{Mirizzi_2009}. The 95\% confidence interval (C.I.) on the coupling constant is at $\mathrm{\leq 6.6 \times 10^{-11} \, GeV^{-1}}$ \cite{2017} from CAST. The RMS fluctuation bounds on non-resonant ALP distortions using Planck data is $ < 18.5 \times 10^{-6}$ \cite{Mukherjee_2019}. Additionally, constraints on the effective number of neutrino species (denoted by $N_{\mathrm{eff}}$) from SO and CMB-S4 can also constrain the couplings of axions or other BSM particles using CMB anisotropies  \cite{dvorkin2022physicslightrelics,https://doi.org/10.17863/cam.30368,green2019messengersearlyuniversecosmic}.
\subsection{{ALP detection using CMB as a backlight}}
\label{sec:microwave}
The CMB is the radiation field from early universe that exhibits tiny temperature fluctuations of the order of $10^{-5}$ K. It behaves like an ideal blackbody with $T_{\mathrm{cmb}} = 2.7255$ K and  intensity given as \cite{Fixsen_2009,fixsen1996cosmic}:
\begin{equation}
\label{eq:cmb}
I_{\mathrm{cmb}} = \left( \frac{2h \nu^3}{c^2}  \right) \frac{1}{e^{h\nu /k_B T_{\mathrm{cmb}}} - 1}.
\end{equation}

CMB cosmology involves the study of its anisotropies (spatial fluctuations in temperature or polarization) \cite{Hu_2002} and spectral distortions (deviations from blackbody spectrum), which are caused due to various photon interactions in early, as well as in late universe, like the $\mu$ and $y$ distortions \cite{Fixsen_1996}, SZ effects \cite{Komatsu_1999,sunyaev1980microwave,1972CoASP...4..173S}, Sachs Wolfe effects \cite{white1996sachs}, lensing \cite{Smith_2007}, etc. The CMB can thus be probed in spectral and spatial domains to look for new physics. The photon-ALP conversion is one such hypothetical interaction that can be probed from these tiny fluctuations in the CMB, if ALPs exist in nature, irrespective of the fact whether they make dark matter components.

The photon-ALP resonant conversion can take place in galaxy clusters in the presence of magnetic fields. This will introduce secondary (late-time) anisotropies in the CMB. This conversion will lead to polarized spectral distortions in the CMB along the cluster line of sight. These conversions can be resonant or non-resonant, with the resonant ones being stronger. The condition for resonance is given by the equality of the effective masses of photon and ALP \cite{Mukherjee_2020}: 
\begin{equation}
\label{eq:resonance cond}
  m_a = m_{\gamma} = \frac{\hbar \omega_p}{c^2} \approx \frac{\hbar}{c^2}\sqrt{n_e e^2 / m_e \epsilon_{0}}, 
\end{equation}
where $\omega_p$ refers to the plasma frequency and $n_e$ is the electron density at the location of conversion. The electron density at the conversion location thus determines the mass of the ALP that is formed. The Lagrangian for the photon-ALP interaction depends on the ALP coupling constant ($g_{a\gamma}$) and the ALP field ($a$), and is given as \cite{Raffelt:1996wa} 
\begin{equation}\label{eq:lagr}
  \mathcal{L}_{\mathrm{int}} = -\frac{g_{a \gamma} F_{\mu \nu}\tilde{F}^{\mu \nu} a}{4} = g_{a \gamma} \vec {E} \cdot \vec {B}_{\mathrm{ext}} a.  
\end{equation}
Here the $\vec {E} \cdot \vec {B}_{\mathrm{ext}}$ term illustrates that the photon polarization parallel to the external magnetic field is involved in the conversion. Thus, the transverse magnetic field along the line of sight is involved in the conversion as an ALP is formed.

ALPs of a particular mass will be formed in a spherical shell within a spherically symmetric cluster, which will be observed as a signal disk along the cluster line of sight. Depending on the electron densities in the cluster, ALPs of masses $10^{-15} - 10^{-11}$ eV can be probed with varying disk sizes for different mass ALPs. The low-mass ALPs will be formed in the outer regions with a larger signal disk, as opposed to the high-mass ALPs, which will have a small signal disk. The probability of conversion at a resonant location is given as \cite{Mukherjee_2020,Mehta:2024wfo} 
\begin{equation}
P_{\mathrm{conv}} = 1 - e^{-\pi \gamma_{\mathrm{ad}}/2}, 
\label{eq:prob}
\end{equation}
with the adiabaticity parameter,
\begin{equation}
\gamma_{\mathrm{ad}} =  \left| \frac{2g_{a\gamma }^2 B_t^2 \nu (1 + z)}{\nabla \omega_p^2} \right|, 
\label{eq:gamma_ad}
\end{equation}
with $B_t$ being the transverse magnetic field, $\nu$ the observing frequency, and $z$ the redshift of the cluster. For a photon travelling through a spherically symmetric galaxy cluster, resonance can take place at two locations within the cluster. The probability of conversion to an ALP is then given by the probability of conversion at one of the two locations, i.e.,
\begin{equation}
    P(\gamma \rightarrow a) = 2(1 - P_{\mathrm{conv}})P_{\mathrm{conv}} = 2e^{-\pi \gamma_{\mathrm{ad}} / 2}(1 - e^{-\pi \gamma_{\mathrm{ad}} / 2}).
   \label{eq: Probab}
\end{equation}
This causes a spectral and spatial change in the CMB along the cluster line of sight, with the intensity change in the non-adiabatic limit ($\gamma_{\mathrm{ad}} << 1$) given as:
\begin{equation}
    \Delta I_{\nu}^{\alpha} = P(\gamma \rightarrow a) I_{\mathrm{cmb}}(\nu) \approx \pi \gamma_{\mathrm{ad}} \left( \frac{2h \nu^3}{c^2}  \right) \frac{1}{e^{h\nu /k_BT_{\mathrm{cmb}}} - 1}.
    \label{eq:Distort}
\end{equation}

We can thus model the ALP signal that will be generated in galaxy clusters, if we have information about their electron densities, magnetic fields and redshifts. These can be inferred from radio observations (Sec.  \ref{sec:radio}), X-rays (Sec.  \ref{sec:xray}) and optical 
surveys (Sec.  \ref{sec:optical}). If clusters are resolvable in all these EM bands (referred to as resolved clusters), we will be able to model the ALP signal from them and can use a pixel-based or power spectrum-based approach \cite{Mehta:2024wfo} to constrain the ALP coupling. If clusters are not resolvable in one or more of these bands, they fall into the domain of unresolved clusters, and will generate a diffused polarized background in the microwave sky, following the distribution of matter density field in the universe \cite{Mehta:2024pdz,mondino2024axioninducedpatchyscreeningcosmic}.

\section{The \texttt{SpectrAx} pipeline} \label{sec:method}

\begin{table}[h!]
\caption{A list of parameters used in estimating the ALPs signal and their significance.}
\label{tab:params}

\begin{tabular}{|c|c|c|}

\hline

\textit{\textbf{Notation}} &
\textit{\textbf{Description}} &
\textit{\textbf{Uniform Prior range}} 

\tabularnewline \hline
$s$ & Magnetic field steepness  &  $0.5 < s < 2$ \\[0.5ex]
 \hline
 $B_0$ & Order of magnetic field at 1 Mpc &  $0.01 \, \mu \mathrm{G \, Mpc}^{s} < B_0 < 0.5 \, \mu \mathrm{G \, Mpc}^{s}$  \\ [0.5ex]
 \hline
 $p$ & Synchrotron spectral index &  $3  < p < 9$  \\ [0.5ex]
 \hline
 $C$ & Synchrotron emission proportionality constant &  $1450  < C < 1550$  \\ [0.5ex]
 \hline
$n_{02}$ & Inner region electron density order &  $\mathrm{5\times 10^{-2} \, cm^{-3}} < n_{02}  < \mathrm{1.5 \times 10^{-1} \, cm^{-3}}$  \\[0.5ex]
 \hline
$n_{0}$ & Outer region electron density order &  $\mathrm{5\times10^{-4}\, cm^{-3}} < n_{0}  < \mathrm{1.5 \times 10^{-3} \, cm^{-3}}$  \\ [0.5ex]
 \hline
$\gamma$ & Transition width parameter & $ 2 < \gamma < 4$  \\[0.5ex]
 \hline
$\alpha$ & Cusp slope parameter &  $1 < \alpha < 3$ \\[0.5ex] 
 \hline
$\beta_1$ & Outer $ \beta$ exponent & $0.5 < \beta_1 < 0.8$  \\ [0.5ex]
\hline
$\beta_2$ & Inner $ \beta$ exponent & $0.8 < \beta_2 < 1.2$ \\ [0.5ex]
 \hline
$r_{s}$ & Scaling radius &  $0.5 \, \mathrm{Mpc} < r_s < 1.5 \, \mathrm{Mpc}$  \\ [0.5ex]
 \hline
$r_{c1}$ & Outer core radius & $\mathrm{0.05 \, Mpc} < r_{c1} < \mathrm{0.7 \, Mpc}$  \\[0.5ex] 
 \hline
$r_{c2}$ & Inner core radius & 
   $\mathrm{0.008 \, Mpc} < r_{c2} < \mathrm{0.05 \, Mpc }$ \\ [0.5ex]
 \hline
$\epsilon$ & Knee slope & $2 < \epsilon < 5$ \\[0.5ex] 
 \hline
 $Z$ & Metallicity &  $ 0.8 < Z < 1.5$\\ [0.5ex]
 \hline
  $A$ & X-ray emission proportionality constant &  $10^{7}  < A < 10^{8}$  \\ [0.5ex]
\hline
$g_{a\gamma}$ & ALP coupling constant & $\mathrm{10^{-14}\, GeV^{-1}} < g_{a\gamma} < 10^{-11} \, \mathrm{GeV^{-1}}$\\ [0.5ex] 
 \hline
\end{tabular} 
\end{table}
The uncertainties surrounding the photon conversion to ALPs include cluster redshift measurements, the transverse magnetic field along the line of sight and the electron density profile of a cluster, all of which 
affect the conversion probability.
 It is not possible to obtain all these quantities unless a multi-probe analysis is done. 

The redshifts are obtained from optical surveys and a cross-matching of the clusters is done for various observations. 
The transverse magnetic field in a galaxy cluster is constrained using synchrotron emission. The electron density and temperature profiles are constrained using thermal X-rays. All these parameters are listed in Table \ref{tab:params} and can be constrained by \texttt{SpectrAx} under the Bayesian model. 
The ALP distortion is also contaminated by the CMB, foregrounds and instrument noise, which require cleaning, to mitigate their impact. All these parts make up pieces of the hierarchical Bayesian framework to obtain constraints on the ALP coupling constant (see Fig. \ref{fig: flowchart}). 
 The uncertainties across various observations are marginalized over to obtain an inference of the coupling strength $g_{a\gamma}$. We apply our pipeline on mock signal to validate the pipeline and benchmark its performance for different multi-band signals.

The parameter estimations are done using the Monte Carlo Markov Chain (MCMC) method using emcee \cite{Foreman_Mackey_2013}. It works on the principle that the true model depends on data and hence, the parameters are estimated using the Bayes' Theorem in which the unnormalized posterior is obtained as the product of the likelihood of data for given parameters values, and the priors on those parameters values:
\begin{equation}\label{eq:bayes}
\mathrm{P(\theta|Data) \propto \mathcal{L}(Data|\theta)\pi(\theta) .}\end{equation}
A Gaussian likelihood has been considered with uniform priors on all parameters within the allowed parameter range in the case of radio and X-ray inference.
During the MCMC parameter estimation, the burn in part of the chain is discarded and thinning is performed to minimize the correlation between steps. The convergence for one of the clusters can be checked in Appendix \ref{sec:converge}.

To demonstrate the working of \texttt{\texttt{SpectrAx}} we perform a mock analysis for 10 mock resolved galaxy clusters, for which we can infer the magnetic field, electron density and redshift from radio, X-rays and optical observations. We consider spherical symmetry for these clusters and assume that the coherence scale of the magnetic field is resolvable, with the high-resolution beam of upcoming detectors. We will find constraints on the ALP coupling constant for ALPs of mass $10^{-14}$ eV (with 15\% tolerance about this mean to set the limits on the conversion length scale as the electron density profile decides the locations where different mass ALPs are formed, as per the resonant condition) from these clusters for a combination of SKA, eROSITA, and SO (or CMB-S4) configuration. Generally, the lower the ALP mass, the stronger is the expected ALP signal, hence better are the constraints.

\subsection{Framework for cross-matching of multi-band observations with optical surveys to obtain the redshifts} \label{sec:cross}
The focus of \texttt{SpectrAx} is to explore clusters resolvable in various multi-frequency observations \cite{csabai2003application,salvato2018finding}. A cluster will have predefined coordinates (ascension and declination) on the 2-dimensional projected sky. If the cluster is well resolved at that sky location for all these bands, the data of the same position from optical surveys can be used to obtain redshift measurements for the resolved cluster. This method filters out clusters that may not be resolvable in one of these bands. The components of the framework corresponding to the band/bands in which these clusters are well resolved can then be used to study them.
Only those clusters that are detectable at all frequencies or contain all the data (magnetic field, electron density, redshifts) will be the ones used for further analysis. Also, the polarized emission from these clusters should be resolvable so that the ALP signal can be constrained.

\begin{figure}[h!]
     \centering
\includegraphics[height=7cm,width=12cm]{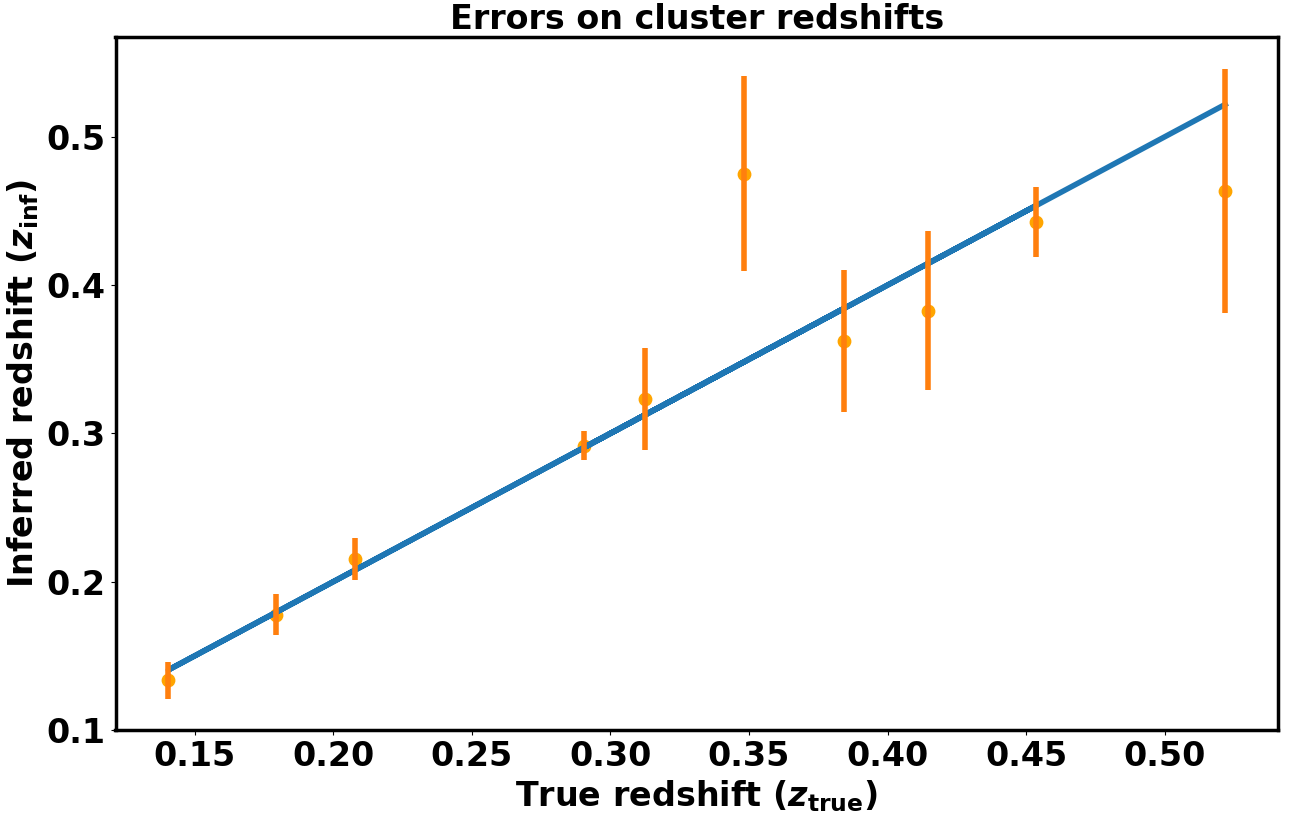}
\caption{Errors on the inferred redshift for the clusters considered. The blue line represents the equality of the redshifts, and the orange dots with error bars represent the errors on the redshifts of the clusters.}
\label{fig: zerr}
\end{figure}

For our mock analysis, we generate ten galaxy clusters in the sky up to a redshift of $z=0.5$, which we consider resolvable. This gives us a probability distribution for the redshift of a cluster. These clusters are assigned random locations in the sky and random redshift values. We assign random relative photo-z errors of up to 10\% on the redshift, i.e., the error on redshifts is:
\begin{equation}
    \sigma(z) = \sigma_z (1 + z), \, \mathrm{with} \, \sigma_z < 0.1 \, . 
\end{equation}
We consider a flat prior on the redshifts in the range of 0.05 to 0.55, and the likelihood is calculated as:
\begin{equation}
\log \mathcal{L} = - \frac{ (z_{\mathrm{true}} -  z_{\mathrm{inf}})^2}{2[\sigma_{z}(1 + z)] ^2 } - 0.5 \, \log [2 \pi \sigma_{z}^2(1 + z)^2] . 
\label{eq:zlike}
\end{equation}
These errors are more than optimum on photometric redshifts and high values are generally obtained for high redshift clusters due to reduced flux. The errors on redshift for the ten galaxy clusters are shown in Fig. \ref{fig: zerr}, with equality being represented by the blue line and the orange dots representing the clusters with the corresponding error bars. Since we have allowed errors with standard deviation ($\sigma_z$) values randomly chosen from a Gaussian distribution, one of the inferred redshifts shows almost a 2$\sigma$ deviation, while one of them at a high redshift, shows a small error bar. 
 
\subsection{Framework for inference of the cluster magnetic field using radio observations}
\label{sec:magnetic}

For the inference of the magnetic field in a cluster, 
the magnetic field profile that has been used is \begin{equation}
  B(r) = B_0 r^{-s}.
\label{eq:mag prof}
\end{equation}
This profile fits well to clusters at low redshifts, based on observations \cite{bonafede2010galaxy}, although a more complex magnetic field profile is needed as per recent studies \cite{osinga2024probing}. \texttt{SpectrAx} is efficient enough to constrain complex profiles as well.

We use the dependence of the synchrotron emission on the magnetic field and spectral index (see Equ.\eqref{eq: pow sync}). Also, a proportionality factor is introduced, which we denote as $C$ and is varied for the clusters to generate signals of about $\sim 0.1$ to 2 $\mu \mathrm{Jy/arcsec^{2}}$  \cite{loi2017magnetic}. The profile parameters ($s$ and $B_0$ in this case), proportionality factor ($C$), and spectral index ($p$) are assigned uniform priors (see Table \ref{tab:params}) and varied within the allowed range for galaxy clusters to generate random profiles for different clusters. We make the synchrotron maps by calculating the synchrotron emission integrated along the line of sight through the cluster for each of these clusters, integrated across the required frequency bands. 

\begin{table}[h!]
\centering
\begin{tabular}{|c|c|c|}

\hline

Bands [GHz] & Noise (in $\mathrm{\mu Jy/beam}$) & Beam (in arcsec)
\tabularnewline \hline

0.77 & 4.4 & 12\\
 \hline
1.4 & 2 & 10\\
 \hline
6.7 & 1.3 & 10\\
 \hline
\end{tabular} 
\caption{Noise and beam used for the SKA bands.}
\label{tab:SKA}
\end{table}

We find constraints for each of these clusters using SKA \cite{loi2017magnetic,heald2020magnetism,braun2019anticipated}. We use the SKA frequency bands at 0.77 GHz, 1.4 GHz, and 6.7 GHz. The noise from the SKA configuration is used.
The noise for these bands are added to the synchrotron maps.

We assign uniform priors to the variables $s$, $p$, $B$ and $C$ within the allowed range.  The Gaussian likelihood used for the inference of magnetic field is 
\begin{equation}
\log \mathcal{L} = - \sum_{\nu}  \left[ \frac{ (\sum_{i = 1}^{n} I_{\mathrm{data}}^i (B_{\perp}^{l}(r_{\perp}),\nu) -  \sum_{i = 1}^{n} I_{\mathrm{mod}}^i (B_{\perp}^{l*}(r_{\perp}),\nu))^2}{2n \sigma_{\nu} ^2 } + 0.5 \, \log [2 \pi n\sigma_{\nu}^2] \right], 
\label{eq:maglike}
\end{equation}
here $I_{\mathrm{data}}^i$ refers to the synchrotron intensity data vector at the pixel marked $i$, $I_{\mathrm{mod}}^i$ refers to the corresponding model intensity, and $\sigma_{\nu}$ is the noise corresponding to the frequency band, mentioned in Table \ref{tab:SKA}. The $B_{\perp}^{l}(r_{\perp})$ refers to the true transverse magnetic field integrated along the line of sight at the sky projected distance $r_{\perp}$ from the cluster center, while $B_{\perp}^{l*} (r_{\perp})$ is the inferred transverse magnetic field integrated along the line of sight. 

\subsubsection*{\textbf{Results on Magnetic field constraints}}
\begin{figure}[h!]
     \centering
\includegraphics[height=11cm,width=14cm]{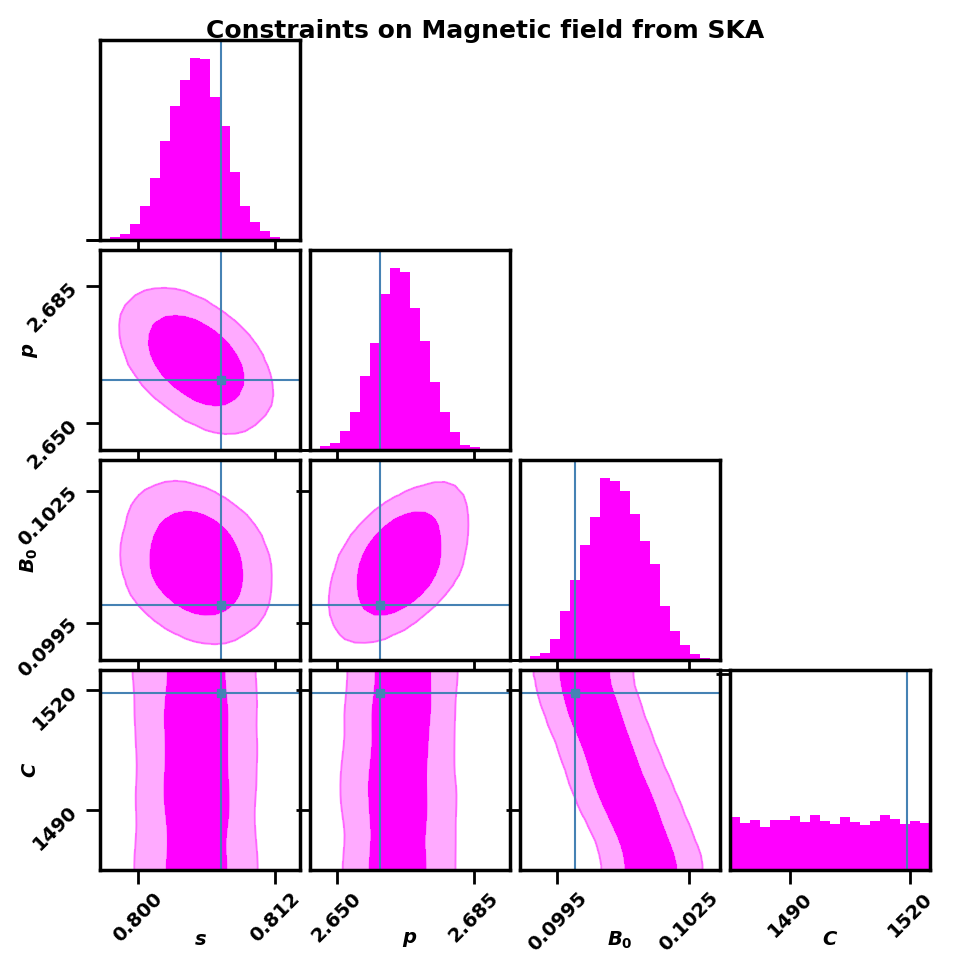}
\caption{Constraints obtained on magnetic field profile parameters ($s$ and $B_0$), spectral index ($p$) and the proportionality factor $C$ for one of the clusters at a redshift $z = 0.32$ using the SKA configuration.}
\label{fig: magcons}
\end{figure}

\begin{figure}[h!]
     \centering
\includegraphics[height=7cm,width=12cm]{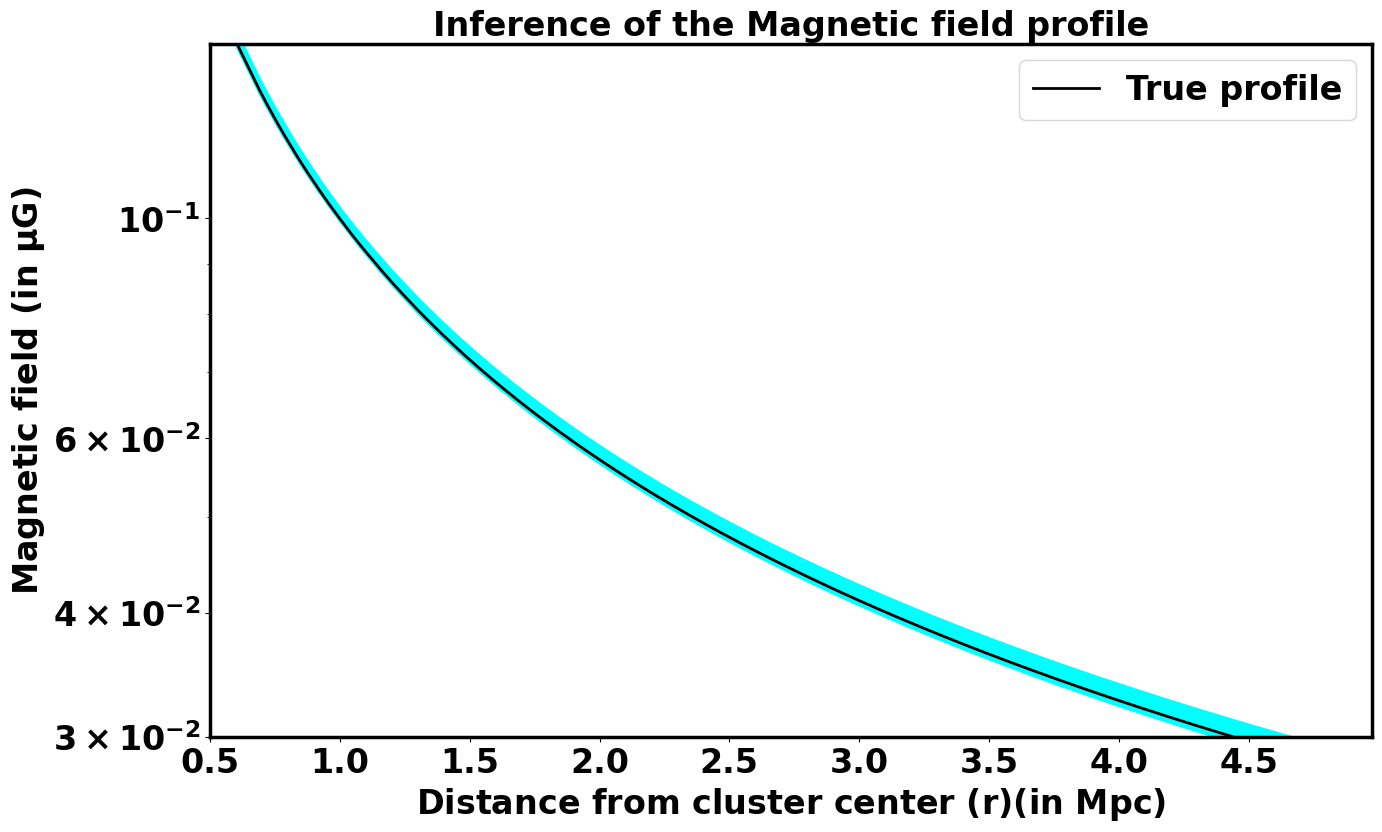}
\caption{The inference of magnetic field for one of the clusters at a redshift of $z = 0.32$, for which the true profile is shown in black line and the 95\% credible interval of the inferred magnetic field in cyan. The angular scale of this cluster on the sky is $\sim 30$ arc-minutes, which using SKA, will be very well resolved with beam sizes of tens of arc-seconds.}
\label{fig: magcons2}
\end{figure}
We show the results on the parameters obtained for one of the clusters in Fig. \ref{fig: magcons}.
The magnetic field $s$ and synchrotron $p$ parameters are relatively well constrained. The parameter $B_0$ exhibits degeneracy with the proportionality factor $C$. The constraints depend on the strength of the signal as well as the number of pixels that the cluster occupies. The average time for the run is $\approx 25$ core-hours per 1000 pixels. As long as the magnetic field coherence angular scale is greater than the beam size, the signal will be resolvable. The 95\% confidence region for the inference of magnetic field for the cluster is shown in Fig. \ref{fig: magcons2}.  With a more complex magnetic field model, the number of parameters will also increase, but \texttt{SpectrAx} can capture those variations at the pixel level. 
The $1/r$ exponent $s$ and magnetic field order $B_0$ will be marginalized over to obtain the ALP coupling constant under the hierarchical Bayesian framework in Sec.  \ref{sec:bayes_axion}.

We have assumed that the magnetic field varies over the angular scale, which is resolvable with the beam resolution of SKA. We have considered a radial profile, but in reality, the clusters will have a different geometry and the profiles will not be radial. This would require complex models with a 3-dimensional variation of the magnetic field. The stochasticity and turbulence in magnetic field will be considered in a future version of \texttt{SpectrAx}. The inference of the magnetic field using synchrotron polarization measurements will provide better constraints on the magnetic field. Using the Faraday rotation $RM$ measurements, the magnetic field profile along the line of sight can also be constrained, which will be able to provide a 3-dimensional inference of the magnetic field direction and magnitude within the cluster. This can be included in an upgraded version of the \texttt{\texttt{SpectrAx}} code.

\subsection{Framework for inference of the cluster electron density using X-rays}\label{sec:electron} 

The electron density profile used is a modified beta model which takes into account the slope at large radii, a cusp core that follows a power law, and the higher electron density in the inner regions \cite{Vikhlinin_2006,mcdonald2013growth,mcdonald2017remarkable,bartalucci2017recovering}: 
\begin{equation}
  n_e^2 = Z\left[ n_0^2 \frac{(r/r_{c1})^{-\alpha}}{(1 + r^2/r_{c1}^2)^{3\beta - \alpha /2}}\frac{1}{(1 + r^{\gamma}/r_s^{\gamma})^{\epsilon / \gamma}} + \frac{n_{02}^2}{(1 + r^2/r_{c2}^2)^{3\beta_2}}\right].
\label{eq:elec dens}\end{equation}
The parameters are defined in Table \ref{tab:params}. An isothermal temperature profile has been used which remains constant until the core radius and then follows a beta model. It is given as \cite{Birkinshaw_1999,de2002temperature} 
\begin{equation}
    T(r)=
    \begin{cases}
        T_{e0} & \text{if } r \leq r_{c1},\\
        T_{e0} \left( 1 + \frac{r - r_{c1}}{r_{c1}}\right)^{-\Gamma }  & \text{if } r > r_{c1}, 
    \end{cases}
    \label{eq:temp profile}
\end{equation}

The X-ray brightness is calculated for the clusters according to Equ.\eqref{eq:xray} using a proportionality factor $A$, which is varied to generate signals of the order of $10^{-13}$ to $10^{-11} \, \mathrm{erg \, cm^{-2} \, s^{-1}}$ \cite{kafer2019toward}. The electron density and temperature profile parameters are varied within the allowed range for different clusters to generate random profiles. The thermal emission X-ray maps are obtained by calculating the line of sight integrated thermal emission from the clusters, across the frequency bands.

We perform the MCMC Bayesian estimation to find constraints on the electron density and temperature profiles for these clusters using the eROSITA specifications \cite{predehl2021erosita}. We use both the soft (0.5 - 2 keV) and hard X-ray (2 - 10 keV) bands, and the corresponding noises are added to the X-ray maps.
{
\begin{table}[h!]
\centering
\caption{Beam and noise used for eROSITA bands}
\label{tab:time}

\begin{tabular}{|c|c|c|}

\hline

Bands & Noise (in $\mathrm{erg \, s^{-1} \, cm^{-2}}$) & Beam (in arcsec)
\tabularnewline \hline

Soft band (0.5 - 2 keV) & $1.1 \times 10^{-13}$ & 28\\
 \hline
Hard band (2 - 10 keV) & $7.1 \times 10^{-13}$ & 40\\
 \hline
\end{tabular} 
\label{tab: eROSITA}
\end{table}
 We assign uniform priors to the variables $n_0$, $r_{c1}$, $\beta_1$, $A$, $T_{e0}$ and $\Gamma$ within the allowed range. The noise from the eROSITA configuration is used. The Gaussian likelihood for the inference of electron density and temperature profiles is given as:
\begin{equation}
\log \mathcal{L} = - \left[ \sum_{\nu} \frac{ (\sum_{i = 1}^{n} I_{\mathrm{data}}^i (n_e^{l}(r_{\perp}),T_e^{l}(r_{\perp}),\nu) -  \sum_{i = 1}^{n} I_{\mathrm{mod}}^i (n_e^{l*}(r_{\perp}),T_e^{l*}(r_{\perp}),\nu))^2}{2n \sigma_{\nu} ^2 } + 0.5 \, \log [2 \pi n\sigma_{\nu}^2] \right],    
\end{equation}
here $I_{\mathrm{data}}^i$ refers to the X-ray intensity data vector at the pixel marked $i$, $I_{\mathrm{mod}}^i$ refers to the corresponding model intensity, and $\sigma_{\nu}$ is the noise corresponding to the frequency band, mentioned in Table \ref{tab: eROSITA}. The $n_e^{l}(r_{\perp})$ and $T_e^{l}(r_{\perp})$ refer to the true electron density and temperature integrated along the line of sight at the sky projected distance $r_{\perp}$ from cluster center, while $n_e^{l*}(r_{\perp})$ and $T_e^{l*}(r_{\perp})$ are the inferred electron density and temperature integrated along the line of sight.}

\subsubsection*{\textbf{Results on Electron density and temperature profile constraints}}

\begin{figure}[h!]
     \centering
\includegraphics[height=13cm,width=16cm]{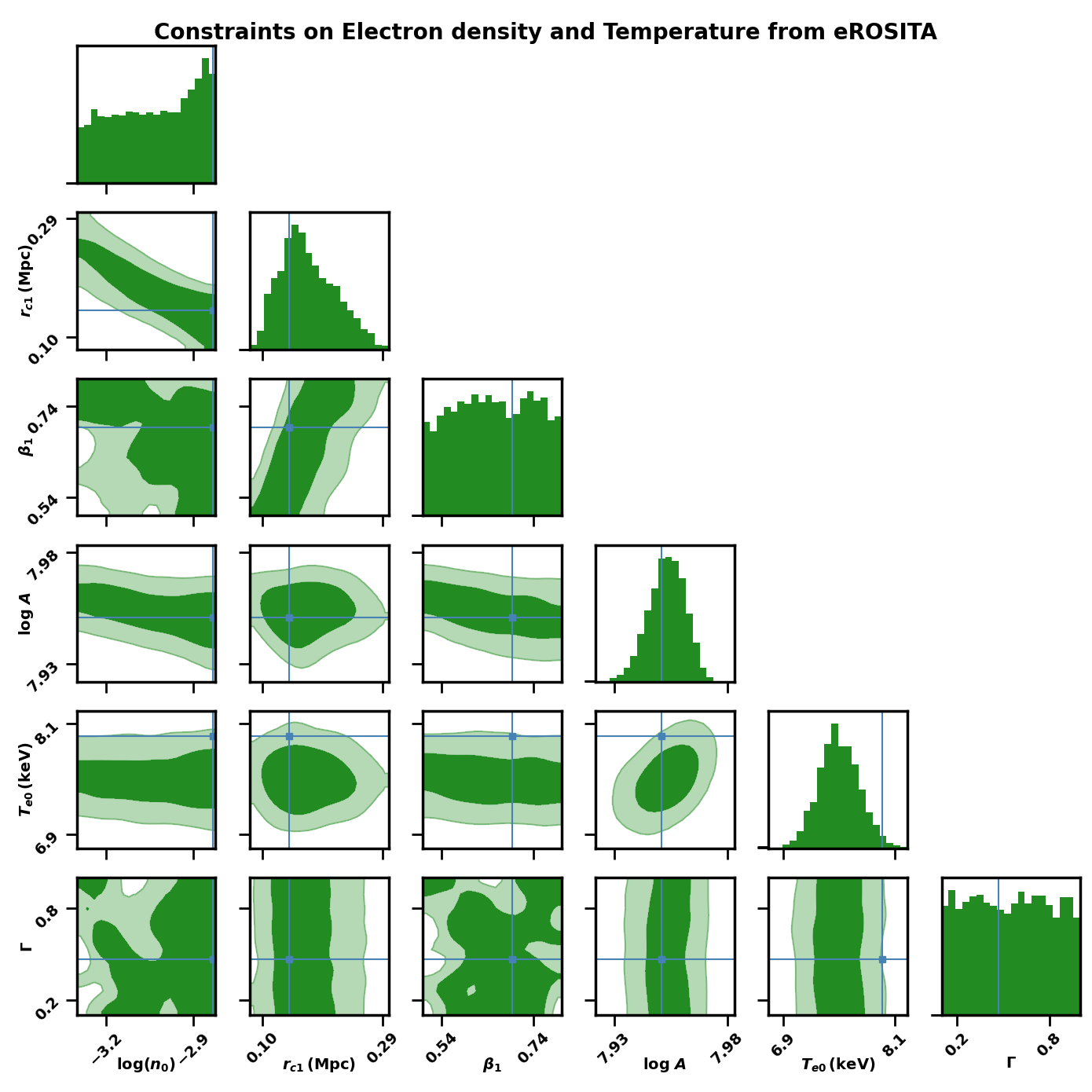}
\caption{Constraints obtained on the parameters of electron density ($n_0$, $r_{c1}$ and $\beta$) and temperature profiles  ($T_e$ and $\Gamma$) of a galaxy cluster at a redshift of z = 0.14, using the eROSITA survey.  }
\label{fig: neTcons}
\end{figure}

\begin{figure}[h!]
     \centering
\includegraphics[height=7cm,width=12cm]{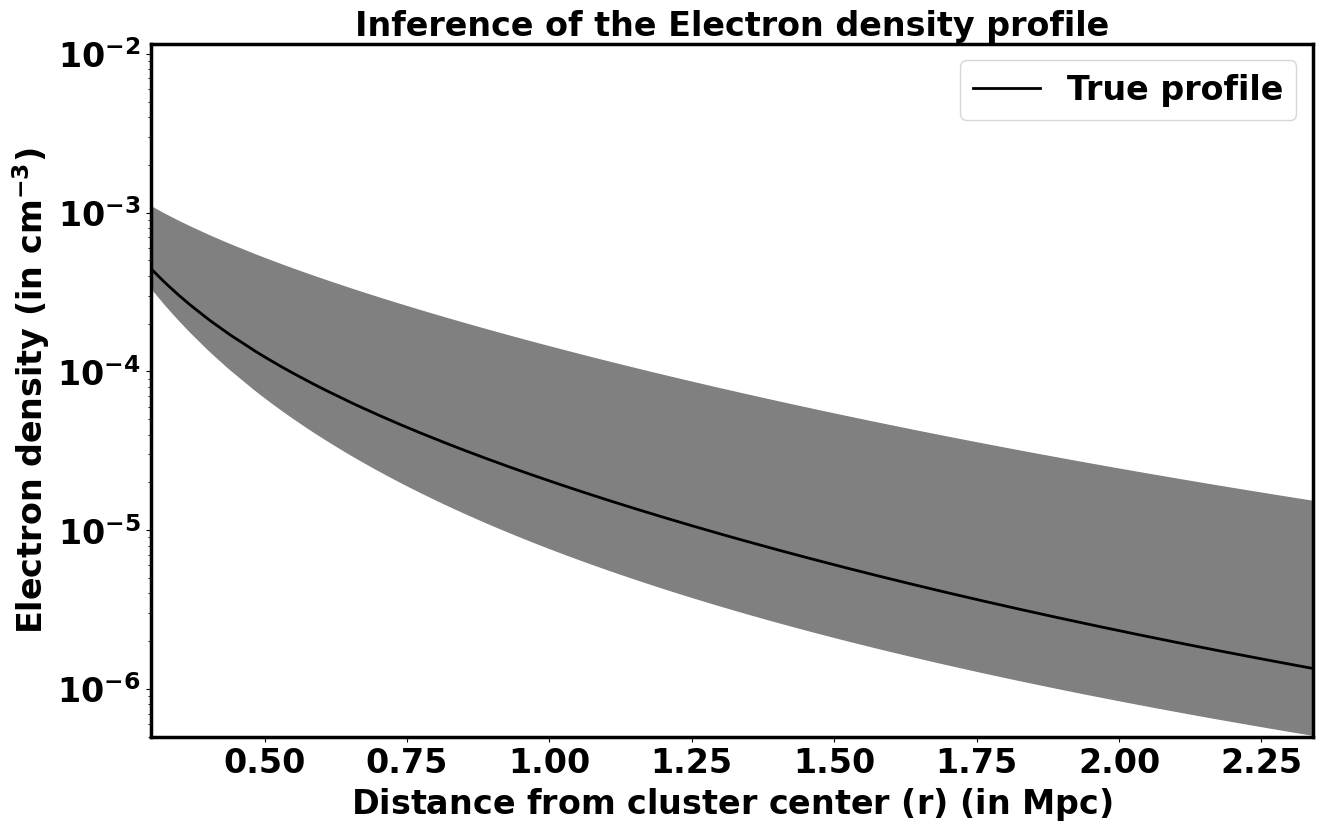}
\caption{The inference of electron density for one of the clusters at a redshift $z = 0.14$, for which the true profile is shown in black line and the 95\% confidence inference of the electron density in grey. The angular scale of this cluster on the sky is $\sim 25$ arcminutes, which using eROSITA, will be very well resolved with beam sizes of tens of arcseconds.}
\label{fig: elecons2}
\end{figure}

The estimation of 6 parameters ($n0$, $r_{c1}$, $\beta_1$, $T_e$, $\Gamma$ and $A$) using X-ray emission is shown is Fig.  \ref{fig: neTcons}. The average time for the runs is $\approx 35$ core-hours per 1000 pixels.
The electron density and temperature profiles involve a large number of parameters and some of them portray degeneracies. The constraints depend on the strength of the signal as well as the number of pixels that the cluster occupies. The 95\% confidence region for the inference of electron density for the cluster is shown in Fig. \ref{fig: elecons2}. 

We have assumed that the electron density varies over the angular scale, which is resolvable with the beam resolution of eROSITA. We have considered a radial profile, but in reality the clusters will have a different geometry and the profiles will not be radial. This would require complex models with 3-dimensional variation of electron density. The turbulence in electron density will be considered in a future version of \texttt{SpectrAx}. \texttt{SpectrAx} works efficiently in estimating a large number of parameters, given computer resources and resolutions are available. The constrained parameters will be marginalized along with the magnetic field parameters and redshift to obtain the ALP coupling constant under the hierarchical Bayesian framework in Sec.   \ref{sec:bayes_axion}.

SpectrAx will be able to employ complex profile models in its code and estimate the parameters. If the profile is simple, accompanied with good-quality data, the estimation will be robust and efficient. If the model is quite complex with a large number of parameters, some of the parameters may portray degeneracies. These degeneracies in profile parameters represent barriers in the way of constraining ALPs, due to astrophysical uncertainties. These can be dealt with by using better and multiple independent cluster observations from various surveys. The three dimensional profile of electron density and magnetic field will be considered in a future version of SpectrAx.

\subsection{ALP signal in CMB map}\label{sec:cmb_axion}
\begin{figure}[h!]
     \centering
\includegraphics[height=11.5cm,width=10cm]{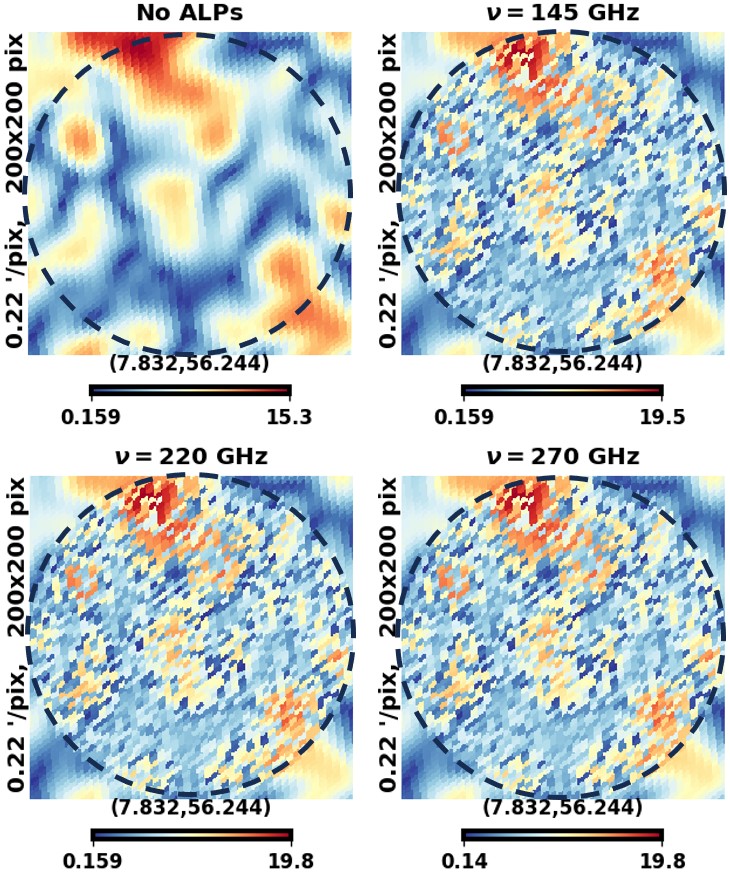}
\caption{This figure depicts the change that is caused in the polarization fluctuations of the CMB at small angular scales (HEALPix NSIDE $=$ 4096) corresponding to the region of $\sim 44 \times 44 $ sq. arcminutes around a resolved cluster in the absence (top left, with CMB being independent of frequency) and presence of ALP signal ($g_{a\gamma} = 5 \times 10^{-12} \, \mathrm{GeV^{-1}}$ and $m_a = 10^{-14} \, \mathrm{eV}$). These fluctuations increase with frequency in the microwave spectrum. The circular region that the cluster occupies shows an increase in fluctuations, while the smooth CMB characteristics are visible in the surrounding region.}
\label{fig: cmbmaps}
\end{figure}

The ALP signal can be probed from polarization measurements of the CMB along the cluster line of sight. As explained in Sec.  \ref{sec:microwave}, the ALP distortion signal for a resolved cluster will be visible as a disk, with its size depending on the mass of the ALPs being formed \cite{Mukherjee_2020,Mehta:2024wfo}. These fluctuations in the cluster will increase with frequency in the microwave spectrum, while the regions surrounding the cluster regions will bear the smooth CMB characteristics, with fluctuations that will be independent of frequency. We show the small-scale anisotropies around a resolved galaxy cluster in different frequency channels accessible from the upcoming CMB missions in Fig.  \ref{fig: cmbmaps}. 

To be able to probe the ALP signal disk, it is necessary that the cluster be resolvable in the CMB. Also, the beam size of the instrument should be smaller than the coherence length of the magnetic field, so that depolarization of the signal does not occur.  We expect that to be the case with the upcoming high-resolution experiments like SO, CMB-S4, etc. Also, the scale of variation of the electron density should be smaller than the scale of the ordered magnetic field. This will let us neglect the effects of turbulence in the magnetic field, which otherwise would lead to partial depolarization of the signal. If multiple mass ALPs are formed, there will be line of sight depolarization depending on the orientation of magnetic field at the conversion locations.

For our mock analysis, we have considered ALPs of masses $10^{-14}$ eV, with a tolerance of $\pm 15 \%$ to be forming in the clusters with a uniform coupling constant of $5 \times 10^{-12} \, \mathrm{GeV^{-1}}$ in the mock data. We consider a uniformly random transverse magnetic field direction. To take this into account, the signal is generated for various clusters at different frequencies assuming a uniformly random polarization direction. Also, the case for zero coupling is considered in Sec. \ref{sec:results} and highlights that the method is unbiased.

The detectability of the ALP signal is still affected by various contaminants.
The CMB is contaminated by foregrounds, such as the galactic synchrotron emission and thermal dust emission. These are highly dominating in the galactic plane and need to be cleaned or accounted for in the analysis. For our mock analysis, we generate the ALP signal in the clusters, obtain the CMB polarization power spectrums from CAMB and the corresponding maps using the synfast routine in HealPy \cite{Zonca2019}, and the foregrounds galactic dust and synchrotron emission from PySM  \cite{Thorne_2017}. We use the dust model "d-3", with a spatially varying spectral index. Also, we use the synchrotron model "s-3" which has a curved spectrum, steepening or flattening with frequency.
The synchrotron emission is strong at frequencies less than $\nu=70$ GHz, while  thermal dust emission impacts the signal for frequencies greater than 
$=200$ GHz. Masking the galactic plane to mitigate their impact thus becomes important, although at a cost of the sky fraction available for analysis.
We do not consider the polarized SZ signals in our analysis as these are weak (of the order of nano-Kelvins) and can be separated based on the distinct spectral and spatial variation of the ALP signal \cite{Mehta:2024wfo,Sazonov_1999,shimon2009power,Louis_2017,Hall_2014,aghanim2008secondary,carlstrom2002cosmology,yasini2016kinetic,challinor2000thermal,sunyaev1980microwave}.

The various mock frequency maps thus contain the following components: the ALP signal, CMB, galactic dust, and synchrotron. These are beam smeared with the instrument beam and the noise corresponding to the frequency is added to the maps. The final map is a combination of different components: CMB ($\mathcal{C}$), ALP distortion ($A$), foregrounds ($S_f$) and noise ($N$). This is given as:
\begin{equation}
S(\nu) = \mathcal{C} + A(\nu) + S_f(\nu) + N(\nu)    ,
\label{eq:data}
\end{equation}

\subsection{Cleaning of the foregrounds}
\label{sec:ilc}
Masking the galactic plane is not enough to remove the foreground contamination. These foregrounds affect the observations even away from the galactic plane. Thus, either their contribution needs to be minimized, or their effect needs to be accounted for. The Interior Linear Combination (ILC) method combines common resolution maps at different frequencies in a manner that minimizes the variance of the combined map, while cleaning the signal \cite{ilc2008internal,Eriksen_2004}. A detailed description of cleaning the CMB maps for ALPs signal is shown in \cite{Mehta:2024wfo}. A weighted linear combination of the various frequency maps is taken based on the covariance of the analysis region ($C_s$) and the spectral variation of the signal.  The higher the number of frequencies, the better the results obtained after ILC.   The weights are obtained as 
\begin{equation}
    w_{\mathrm{ilc }} =  f_{a \gamma}^T C_s^{-1}(f_{a \gamma}^T C_s^{-1} f_{a \gamma})^{-1}    
    ,
    \label{eq:ilc w}
\end{equation}
where $f_{a\gamma}$ is the ALP distortion spectrum dependence which at the intensity level goes as  $f_{a\gamma} \propto \nu I_{\mathrm{cmb}}(\nu)$. The ILC cleaned map is given by
\begin{equation}
    S_{\mathrm{ilc}} = 
 \sum_{\nu} w_{\mathrm{ilc}}^{\nu}  S(\nu). 
\end{equation}

\texttt{SpectrAx} is also capable of employing a template matching of foregrounds by considering the spectral variation of foregrounds at the frequencies where they are dominant (below 70 GHz for synchrotron and above 200 GHz for dust), based on an assumed spectral variation model \cite{Hansen_2006}. This spectral variation is then used to estimate the contribution of the foregrounds at the matching frequencies between 90 and 160 GHz, where the foreground contamination is low.  
 This technique enables us to probe the signal at a higher resolution as compared to ILC, but suffers from bias effects in case the foreground contamination is not well modelled. We present results for the ILC cleaning in this analysis. The implementation of template matching can be found in our previous works \cite{Mehta:2024wfo,Mehta:2024pdz}. 

 We present the results obtained using the configurations of the SO \cite{Ade_2019} and CMB-S4 \cite{abazajian2016cmbs4} detectors. The SO
will observe in 6 bands from 27 to 280 GHz. CMB-S4 will use 6 observing bands from 30 to 270 GHz as per the current plan. These will be able to map tens of thousands of galaxy clusters up to a redshift of $z=3$. Also, they will be able to detect polarization fluctuations of the order of tens of nano-Kelvins.

\begin{table}[htbp]
  \small
  \centering
  \caption{\textbf{Instrument noise and beams for SO and CMB-S4 detectors.} }
  \subfloat[\textbf{Simons observatory}]{%
    \hspace{0.5cm}%
    \begin{tabular}{|c|c|c|}
        \hline
        $\nu$ (in GHz) & Noise (in $\mu$K) & Beam \\
        \hline
        \hline
        27 & 52 & 7.4 \\
        \hline
        39 & 27 & 5.1 \\
        \hline
        93 & 5.8 & 2.2 \\
        \hline
        145 & 6.3 & 1.4 \\
        \hline
        225 & 15 & 1.0 \\
        \hline
        280 & 37 & 0.9 \\
        \hline
    \end{tabular}%
    
    \hspace{.5cm}%
  }\hspace{1cm}
  \subfloat[\textbf{CMB-S4}]{%
    \hspace{0.5cm}%
    \begin{tabular}{|c|c|c|}
        \hline
        $\nu$ (in GHz) & Noise (in $\mu$K) & Beam \\
        \hline
        \hline
        30 & 30.8 & 7.4 \\
        \hline
        40 & 17.6 & 5.1 \\
        \hline
        95 & 2.9 & 2.2 \\
        \hline
        145 & 2.8 & 1.4 \\
        \hline
        220 & 9.8 & 1.0 \\
        \hline
        270 & 23.6 & 0.9 \\
        \hline 
    \end{tabular}%
    \hspace{.5cm}%
  }\label{tab:noise}
\end{table}
The contribution of galactic synchrotron is weak above 70 GHz and is dominated by other signals (CMB, dust, etc.). This lets us use the frequency bands with higher resolution (above 90 GHz) for our analysis. Thus the maps are smoothed by the common beam of 2.2 arcminutes corresponding to the frequency bands 93 GHz for SO and 95 GHz for CMB-S4 with the corresponding instrument noises (see Table \ref{tab:noise}). The ILC is performed around the cluster region to obtain the ILC-cleaned map. 

\subsection{Multi-band Bayesian framework to search for ALPs}\label{sec:bayes_axion}
The polarization information to estimate the ALP signal at the level of Q and U maps is not known \cite{2020_planck,Austermann_2012}. Thus, we will have to constrain the coupling constant using the intensity map ($I_{\mathrm{pol}}$), where,
\begin{equation}
    I_{\mathrm{pol}} = \sqrt{I_{Q}^2 + I_{U}^2}.
\end{equation}

\texttt{SpectrAx} can also probe the ALP signal at the power spectrum level, but that requires precise measurements of the electron densities and magnetic field profiles of galaxy clusters. We have shown the power spectrum-based approach in \cite{Mehta:2024wfo}. Here we probe the signal at the pixel level, by marginalizing over the nuisance parameters.  

The coupling constant determination involves the multi-band hierarchical Bayesian analysis. This technique involves marginalizing the nuisance parameters to obtain posteriors for required parameters. The posteriors for the magnetic field and electron density profile parameters are marginalized over to obtain the posterior on the coupling constant $g_{a \gamma}$ using the ILC cleaned map. The error in redshift is obtained from the optical surveys.  The errors on the magnetic field, electron density, and redshifts get propagated to provide constraints on the coupling constant $g_{a\gamma}$.  A uniform prior  on $g_{a\gamma}$   is used for values from  $\mathrm{10^{-14}}$ to $\mathrm{1.5 \times 10^{-11} \, GeV^{-1}}$ in this analysis, with an increased range from Table \ref{tab:params}, to check for the stability of the results in case of variation in prior range. 
 This is given as a $p + q + 1$ order integration, with $p$ being the number of magnetic field parameters, $q$ being the number of electron density parameters that are being marginalized over and $z$, the redshift distribution 
\begin{equation}
\label{eq:pgag}
   P(g_{a\gamma} | \mathrm{Data}) \propto \pi (g_{a\gamma} ) \int \int \int \mathcal{L}(\mathrm{Data} | g_{a\gamma},\{B \},\{n_e\},z) P(\{B\})P(\{n_e\}) \mathrm{d}^p\{B\} \mathrm{d}^q\{n_e\} \mathrm{d}z.
\end{equation} 
We use the posteriors obtained from radio (see Sec.  \ref{sec:magnetic}) and X-ray (see Sec.  \ref{sec:electron}) surveys as priors for the nuisance parameters. The simultaneous Bayesian estimation of the nuisance parameters and ALP coupling constant is performed.
The Gaussian likelihood used for $n$ pixels in a cluster is: 
\begin{equation}
\log \mathcal{L} = - \frac{ (\sum_{i = 1}^{n} I_{\mathrm{data}}^i)^2 - (\sum_{i = 1}^{n} I_{\mathrm{mod}}^i)^2 - n\langle \sigma \rangle^2}{2 \, \mathrm{Cov(ax,\sigma)}} - 0.5 \log[2\pi \,  \mathrm{Cov(ax,\sigma)} ],
\label{eq:axlike}
\end{equation}
here $I^{i}$ refers to the intensity at a pixel numbered $i$, $I_{\mathrm{data}}^{i}$ refers to the intensity corresponding to the particular pixel $i$ in the ILC cleaned mock data map, $I_{\mathrm{mod}}^{i}$ refers to the modelled ALP distortion intensity corresponding to the particular pixel $i$, and $\mathrm{Cov(ax,\sigma)}$ denotes the covariance matrix given by
\begin{equation}
\mathrm{Cov(ax,\sigma)} = n \langle \sigma \rangle^2 + 2 \sum_{i = 1}^{n} \langle Q_{\mathrm{ax}}^i \cdot Q_{\mathrm{\sigma}}^i + U_{\mathrm{ax}}^i \cdot U_{\mathrm{\sigma}}^i \rangle, 
\label{eq:axcov}
\end{equation}
where 
$\langle \sigma \rangle$ refers to the standard deviation obtained from various realizations of non-ALP fiducial maps, $Q$ and $U$ represent the two linear polarizations from different realizations of the ALP signal (denoted as $\mathrm{ax}$) and realized noise (denoted as $\mathrm{\sigma}$). 

In the absence of an ALPs signal, the terms associated with $Q_{\mathrm{ax}}$ and $U_{\mathrm{ax}}$ are zero. However, in the presence of the ALPs signal, the second term in Equ.\eqref{eq:axcov} need not vanish after summing over many pixels, if the number of pixels is less. So, we need to calculate this number for an accurate estimation of the coupling constant. For a pessimistic choice of $g_{a \gamma}=10^{-11}$ GeV$^{-1}$, close to the current bound from CAST \cite{2017}, we calculate the covariance matrix. If we consider coupling strength to be zero in the covariance matrix, then the uncertainty in the inference from the CMB map would reduce, but could also lead to biased results if ALPs exist with a non-zero coupling constant. In the \texttt{\texttt{SpectrAx}} code, the covariance matrix can be calculated for both zero and non-zero values of the coupling constant. The covariance will be estimated using different realizations of the ALP signal, calculated with a high value of the coupling constant ($10^{-11} \, \mathrm{GeV^{-1}}$ in this analysis) and varying electron density and magnetic field profiles. The use of a high coupling constant will give us conservative constraints on the coupling constant, as it increases the covariance. The covariance for a cluster region due to various components (CMB, ALP signal, foregrounds, instrument noise, etc.) is obtained as the ensemble average of the variance of multiple cluster-sized regions (with different profiles and ALP signals for a high coupling constant) in the ILC cleaned map.

For constraining the coupling constants from resolved clusters, separate mass windows can be used as different mass conversions occur at different distances from the cluster center. This changes the size of the signal disk, which is brighter in outer regions of the cluster where low mass ALPs can form \cite{Mehta:2024wfo}.

\section{Results}
\label{sec:results}
We present the constraints obtained on the coupling constant using the hierarchical Bayesian framework for different sets of multi-band surveys for two scenarios of coupling strengths, (i) $g_{a\gamma} = 0$ and (ii) $g_{a\gamma} = 5\times 10^{-12}$ GeV$^{-1}$. We have the magnetic field constraints from our mock analysis using SKA configuration, electron density constraints from eROSITA configuration, and CMB observations from SO and CMB-S4.  

\begin{figure}[h!]
     \centering
\includegraphics[height=7cm,width=12cm]{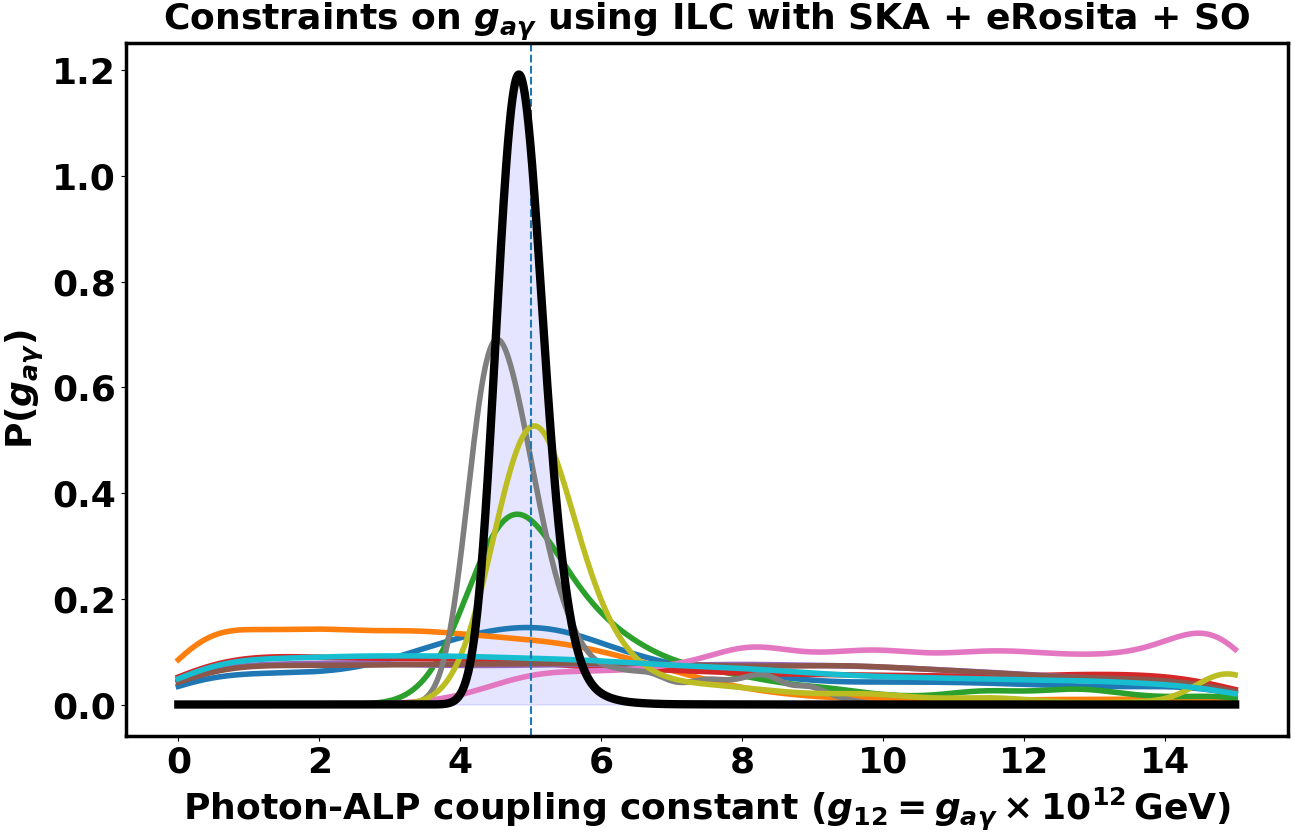}
\caption{Constraints on $g_{a\gamma}$ ($m_a = 10^{-14}$ eV) for a combination of SKA, eROSITA, and SO for the 10 clusters. The bold black line represents the merged posterior. True value is shown as a dashed vertical line, which is $g_{a\gamma} = 5 \times 10^{-12} \, \mathrm{GeV}^{-1}$.}\label{fig:combo_1}
\end{figure}

\begin{figure}[h!]
     \centering
\includegraphics[height=7cm,width=12cm]{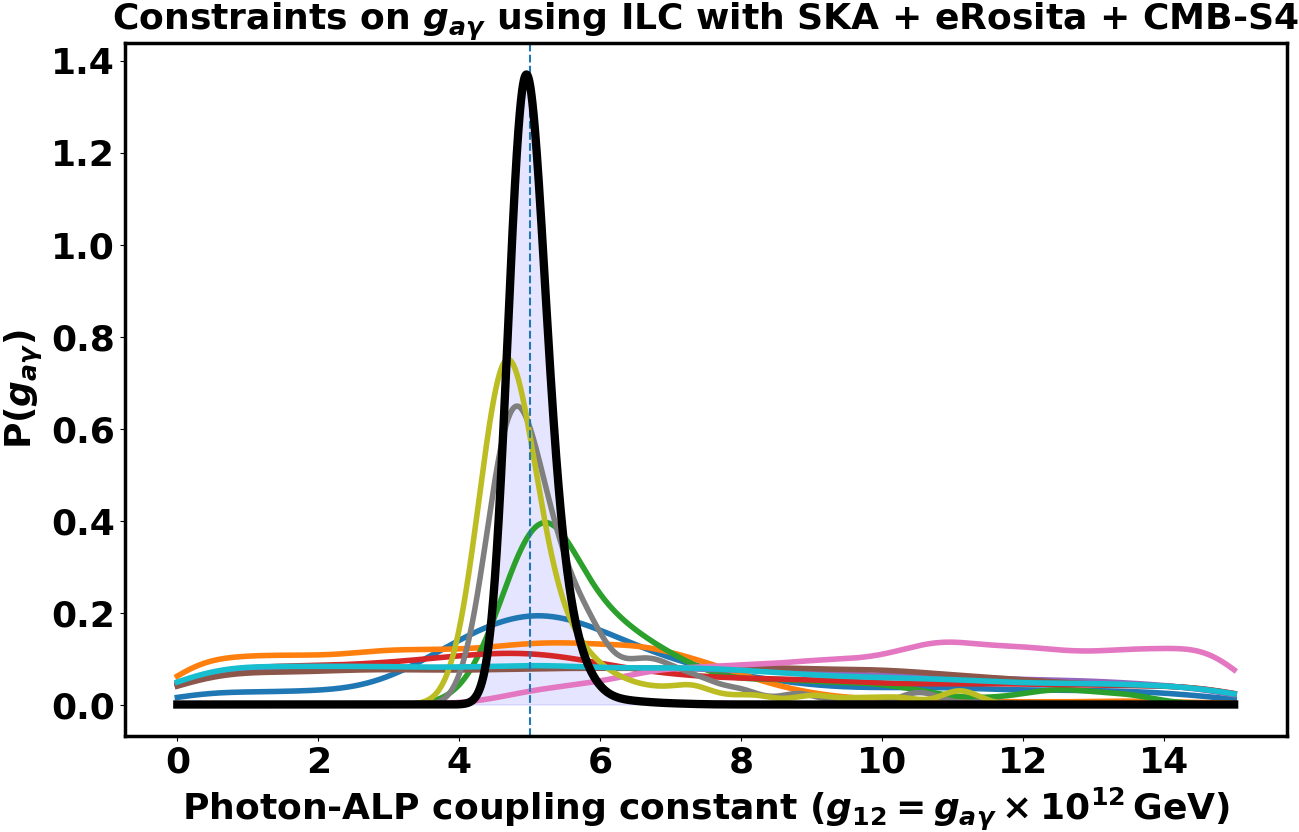}
\caption{Constraints on $g_{a\gamma}$ ($m_a = 10^{-14}$ eV) for a combination of SKA, eROSITA and CMB-S4 for the 10 clusters shown in different colours. The bold black line represents the merged posterior. True value is shown as a dashed vertical line, which is $g_{a\gamma} = 5 \times 10^{-12} \, \mathrm{GeV}^{-1}$.}\label{fig:combo_3}
\end{figure}

\subsection{Zero ALP coupling ($g_{a\gamma} = 0$):}
\label{sec:zero}

\begin{figure}[h!] 
\centering
\includegraphics[height=7cm,width=12cm]{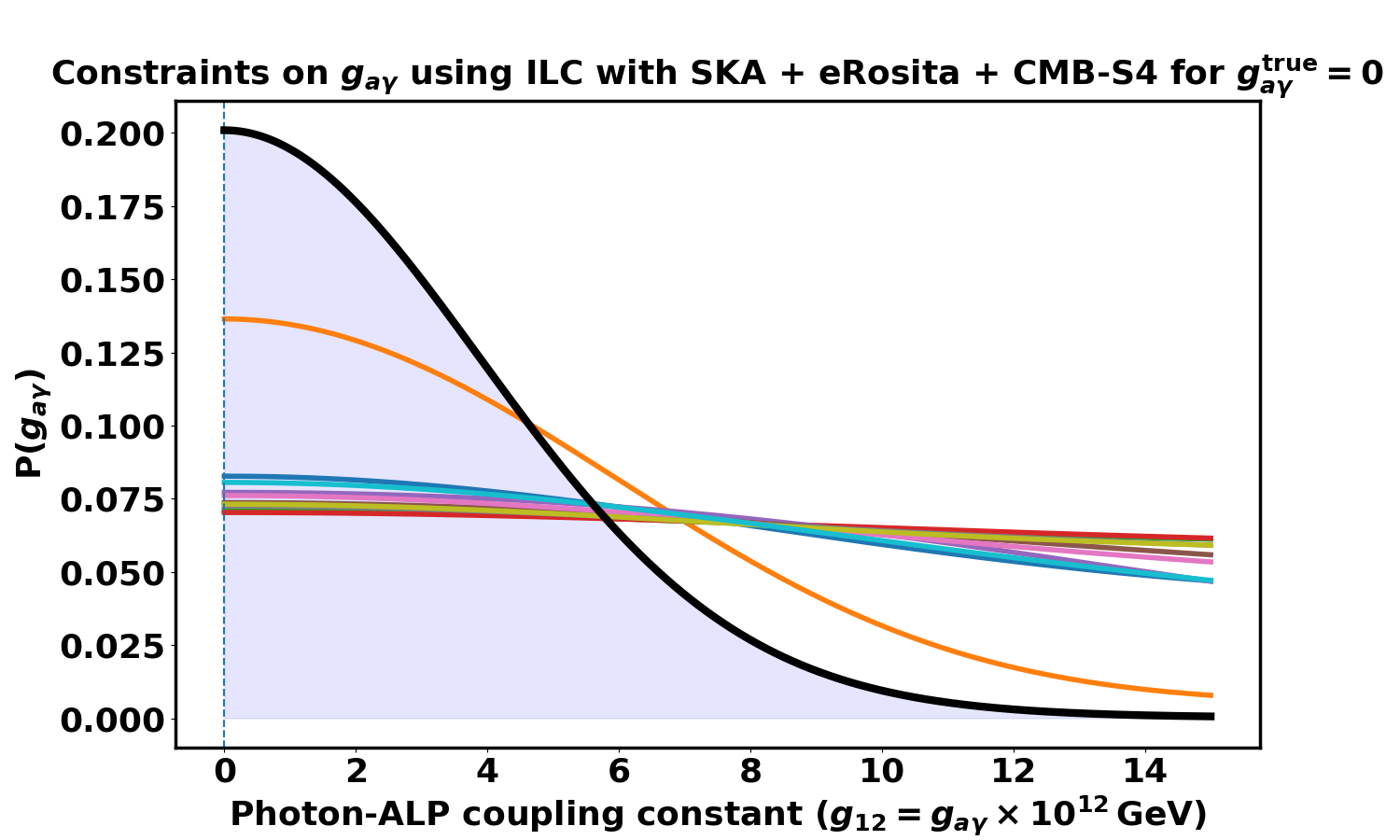}
\caption{Constraints on $g_{a\gamma}$ for a combination of SKA, eROSITA and CMB-S4. The curves represent the constraints obtained for the various clusters when the true value in mock data is $g_{a\gamma} = 0$. The true value is shown as a dashed vertical line.} 
\label{fig:zeros}
    
\end{figure}
\texttt{SpectrAx} gives reliable results in the case of no coupling ($g_{a\gamma} = 0$) as well. This case will give an upper bound on the photon-ALP coupling constant, and can only be further probed with higher resolution and sensitive detectors.

 Here we check the reliability of \texttt{SpectrAx} in case of non-existence of ALPs ($g_{a\gamma} = 0$). We present the results of zero coupling for the clusters in Fig. \ref{fig:zeros}. These constraints are obtained using a combination of SKA $+$ eROSITA $+$ CMB-S4. The constraints for these clusters using the same configurations for a coupling constant of $g_{a\gamma} = 5 \times 10^{-12} \, \mathrm{GeV^{-1}}$ can be seen in Fig. \ref{fig:combo_3}. 

Currently the bounds on coupling constant from CAST are at $g_{a\gamma} = 6.6 \times 10^{-11} \, \mathrm{GeV}^{-1}$. The upcoming detectors will be observing tens of thousands of galaxy clusters. Constraints from thousands of clusters will provide much better bounds on the ALP coupling constant.
Similar analysis at the power spectrum level can also be performed for resolved as well as unresolved galaxy clusters. The bounds that can be obtained from a large number of clusters accessible with SO and CMB-S4 for the case of no-ALPs are studied in \cite{Mehta:2024wfo} (for the case of $10^{-13}$ eV mass ALPs) and \cite{Mehta:2024pdz}. We are able to obtain bounds on the ALP coupling constant, up to 95\% confidence interval for ALP mass $10^{-13}$ eV at $g_{a\gamma} < 5.2 \times 10^{-12} \, \mathrm{GeV^{-1}}$ using clusters up to redshift $z = 1$ with SO (which will observe 24000 clusters upto $z \sim 2$), and $g_{a\gamma} < 3.6 \times 10^{-12} \, \mathrm{GeV^{-1}}$  with CMB-S4 (which will observe 70000 clusters up to $z \sim 2$).

\subsection{Non-zero ALP coupling ($g_{a\gamma} = 5 \times 10^{-12} \, \mathrm{GeV^{-1}}$)}
\texttt{SpectrAx} gives an accurate inference of the ALPs coupling strength for various ALP masses. Here we analyze the case for $10^{-14}$ eV ALP mass. We show the constraints using a combination of SKA $+$ eROSITA $+$
 SO in Fig. \ref{fig:combo_1} and SKA $+$ eROSITA $+$
 CMB-S4 in Fig. \ref{fig:combo_3}. The various lines represent constraints from different clusters, with the bold black line in each plot obtained by merging constraints from all clusters, by multiplying their posteriors. The shaded region represents the acceptable values for the ALP coupling constant. It can provide constraints on the coupling constant in $\sim 50$ core-hours per 1000 pixels.
 
 The best results are obtained for a combination of SKA, eROSITA and CMB-S4 surveys, owing to a higher sensitivity of CMB-S4. 
A significant feature is the lower bound on the coupling constant that may be obtained for a non-zero value of the coupling constant, if ALPs exist. The results improve with higher sensitivity and resolution of the surveys.
The constraints for low-mass ALPs will be better than the constraints for high-mass ALPs as the signal strength is higher for low mass ALPs. 
We can also improve the constraints by using the SZ effect or lensing mass measurements from galaxy clusters by taking into account the variation in the cluster properties, based on the cluster mass. This we plan to include in the upgraded version of the framework \texttt{\texttt{SpectrAx}}. The template matching of foregrounds can also provide tighter constraints at a higher resolution on the coupling constant as well, but suffers from bias if the foregrounds are not modelled well. If the polarization information is known along the line of sight, we can get tighter constraints at the polarization maps level.

\section{Conclusion} \label{sec:conclusion}

The multi-band framework \texttt{\texttt{SpectrAx}} brings together the astrophysical and cosmological aspects of galaxy clusters under one umbrella. By combining information from different EM bands, it can make a data-driven Bayesian inference of the ALP signal. It will provide the basis to study the properties of clusters such as their magnetic fields, temperature, and electron densities, while also being able to combine them to gain insight into cosmological phenomena to explore new physics such as ALPs. We have demonstrated the performance of this method on simulated mock data.

The transverse magnetic field profile in galaxy clusters can be obtained from \texttt{SpectrAx} using the synchrotron emission (see Sec.   \ref{sec:magnetic}). This emission peaks in the radio spectrum and experiments like SKA will be able to constrain the magnetic field in clusters.
The effects of turbulent magnetic fields will require complex magnetic field models, but \texttt{SpectrAx} is efficient in constraining them using its pixel-based Bayesian approach. 

The thermal emission from galaxy clusters can be used to obtain their electron density and temperature profiles using X-rays. The eROSITA currently stands the best X-ray telescope in terms of sensitivity and resolution. Using \texttt{SpectrAx}, we will be able to infer these profiles of clusters from the eROSITA survey (see Sec.  \ref{sec:electron}). The pixel-based approach of \texttt{SpectrAx} can constrain complex profiles as well.

If galaxy clusters are resolvable in different EM bands, \texttt{SpectrAx} can infer their magnetic field from radio synchrotron emission, and electron density and temperature from X-ray emission. Using redshift measurements from optical surveys \ref{sec:cross}, \texttt{SpectrAx} will be able to obtain constraints on the ALP coupling constant $g_{a\gamma}$ under a hierarchical Bayesian framework through the CMB polarization observations from detectors like SO and CMB-S4 (see Sec.  \ref{sec:bayes_axion}). \texttt{SpectrAx} is able to infer the ALP coupling constant using 6 nuisance parameters in $\approx 50$ core-hours per 1000 pixels (see Table \ref{tab:time}). SpectrAx uses parallelization in a way that can constrain the parameters for multiple galaxy clusters together, thus data from multiple clusters can be analyzed alongside. This enables an efficient use of multiple core machines and saves time.
All the values quoted above are model and parameters dependent, but \texttt{SpectrAx} performs at an optimum pace and efficiency in being able to constrain complex models. 
\begin{table}[h!]
\centering
\caption{Analysis case and average time required for constraints (CPU clock speed = 3.5 GHz).}
\label{tab:time}

\begin{tabular}{|c|c|}

\hline

\textit{\textbf{Analysis case}} &
\textit{\textbf{Average time required}}
\tabularnewline \hline

Magnetic field inference & 25 core-hours/1000 pixels \\
 \hline
Electron density \& Temperature inference & 35 core-hours/1000 pixels  \\
 \hline
ALPs coupling constant inference & 50 core-hours/1000 pixels  \\
 \hline
\end{tabular} 
\end{table}

\texttt{SpectrAx} can also perform a template matching of foregrounds to take into account the foreground contamination, based on an assumed spectral shape of the foregrounds. This will enable the probing of ALP coupling at a higher resolution, but at the cost of residual bias in case of inefficient modelling of foregrounds. Also, it can apply a power spectrum-based approach using precise measurements of electron density, magnetic field, and redshift to constrain the ALP coupling constant as has been shown in our previous works \cite{Mehta:2024pdz,Mehta:2024wfo}. The strength of the ALP signal depends on the ALP mass and the coupling constant. Thus the ALP mass and coupling relation can be constrained. 

There are also other astrophysical uncertainties that can affect the ALP signal. The ellipticity and geometry of clusters will impact the shape of the signal disk. Also, the turbulent magnetic field will cause a partial depolarization of the signal when the magnetic field coherence angular scale is smaller than the beam size. These will be considered in a future analysis. The mass measurements (from lensing, or X-rays and optical surveys) of clusters can also be used to obtain constraints on the electron density.
Also, SZ effects can be used to obtain constraints on the electron density and temperature profiles of clusters \cite{Birkinshaw_1999}. These will be considered in a future work of upgraded \texttt{\texttt{SpectrAx}}. If ALPs exist, there will also be a background ALP signal from unresolved clusters. These clusters are not resolvable in different EM bands, but the ALP diffused signal can be probed via the power spectrum approach \cite{Mehta:2024pdz}. This signal will also be correlated with the large-scale structure \cite{mondino2024axioninducedpatchyscreeningcosmic}. The current setup of \texttt{\texttt{SpectrAx}} does not explore this effect.

In the near future, the number of galaxy clusters detectable with the ongoing/upcoming surveys in radio, CMB, optical, and X-ray bands, will be a few tens of thousands with much higher angular resolutions than now. \texttt{\texttt{SpectrAx}} can constrain radial profiles of magnetic field and electron density using these observations, along with finding beyond standard model physics. The pipeline will be made publicly available for ALPs analysis from CMB data after its demonstration on data from CMB experiments such as the SPTpol \cite{Austermann_2012} and ACTpol \cite{niemack2010actpol,thornton2016atacama}. The method developed in this framework will pave the way for astrophysical and cosmological studies in this era of precision cosmology in the coming years.
 \appendix
\section{Convergence of MCMC chains for the inference of ALPs coupling constant}
\label{sec:converge}

\begin{figure}[h!] 
\centering
\includegraphics[height=7.5cm,width=15cm]{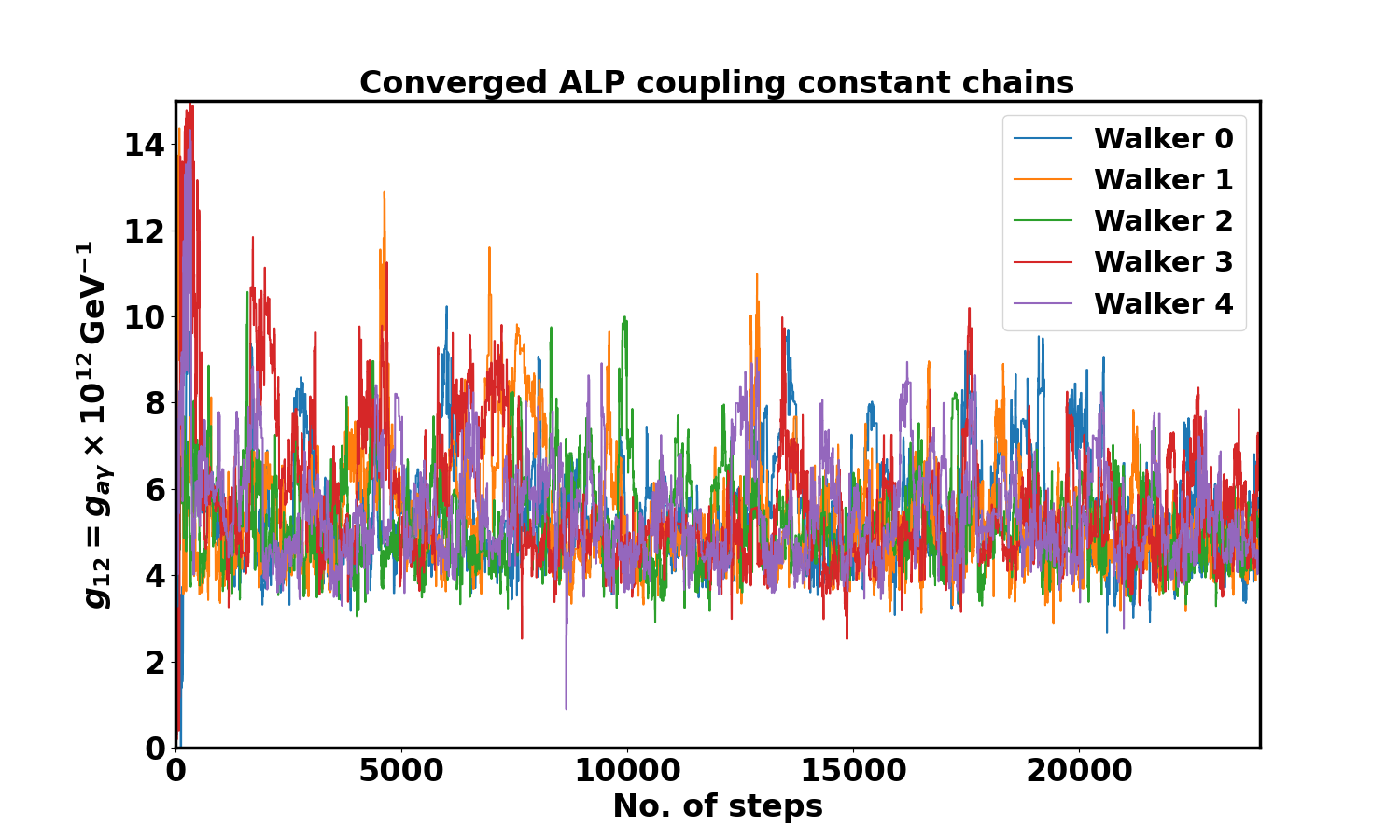}
\caption{We show five randomly chosen walkers (out of 25 walkers) for the ALPs coupling constant $g_{a\gamma}$ as a function of the MCMC steps obtained using \texttt{SpectrAx} for a single galaxy cluster.} 
\label{fig:gaconv}
    
\end{figure}

We show the convergence of MCMC chains for one of the galaxy clusters in Fig. \ref{fig:gaconv}. Here we have considered 25 walkers (in the figure we show 5 of them), which are represented by different colours. The Gelman-Rubin statistic for this case is 0.99 . The variation in the estimation of nuisance magnetic field and electron density parameters (which are marginalized over) leads to the high deviations from the true value when the number of steps is low. The chains converge to the true value of $g_{a\gamma} = 5 \times 10^{-12} \, \mathrm{GeV^{-1}}$ when the number of steps increases, accompanied with a decrease in departure of the walkers from the true value. Thus, the earlier part of the chains needs to be discarded. Moreover, thinning is required in the rest of the chain to keep the samples independent.
The thinning is performed for all clusters at about 20 steps after discarding 18000 steps. We perform these operations to obtain the posteriors on the ALP coupling constant from different clusters as shown in Sec. \ref{sec:results}.

\acknowledgments
    This work is a part of the $\langle \texttt{data|theory}\rangle$ \texttt{Universe-Lab}, supported by the TIFR  and the Department of Atomic Energy, Government of India. The authors express their gratitude to the TIFR CCHPC facility for meeting the computational needs. Furthermore, we would also like to thank the SKA, eROSITA, CMB-S4, and Simons Observatory collaborations for providing the instrument noise and beam resolutions. 
 Also, the following packages were used for this work: Astropy \cite{astropy:2013,astropy:2022,astropy:2018},
, NumPy \cite{harris2020array}
CAMB \cite{2011ascl.soft02026L}, SciPy \cite{2020SciPy-NMeth}, SymPy \cite{10.7717/peerj-cs.103}, Matplotlib \cite{Hunter:2007}, emcee \cite{Foreman_Mackey_2013}, HEALPix (Hierarchical Equal Area isoLatitude Pixelation of a sphere)\footnote{Link to the HEALPix website http://healpix.sf.net}\cite{2005ApJ...622..759G,Zonca2019} and PySM \cite{Thorne_2017}.

\bibliography{references.bib}

\providecommand{\href}[2]{#2}\begingroup\raggedright\begin{thebibliography}{100}

\bibitem{alpher1948evolution}
R.A.~Alpher and R.~Herman, \emph{Evolution of the universe}, {\emph{Nature} {\bfseries 162} (1948) 774}.

\bibitem{ratra2008beginning}
B.~Ratra and M.S.~Vogeley, \emph{The beginning and evolution of the universe}, {\emph{Publications of the Astronomical Society of the Pacific} {\bfseries 120} (2008) 235}.

\bibitem{gamow1948evolution}
G.~Gamow, \emph{The evolution of the universe}, {\emph{Nature} {\bfseries 162} (1948) 680}.

\bibitem{Dodelson:2003ft}
S.~Dodelson, \emph{{Modern Cosmology}}, Academic Press, Amsterdam (2003).

\bibitem{brunetti2014cosmic}
G.~Brunetti and T.W.~Jones, \emph{Cosmic rays in galaxy clusters and their nonthermal emission}, {\emph{International Journal of Modern Physics D} {\bfseries 23} (2014) 1430007}.

\bibitem{bonafede2009revealing}
A.~Bonafede, L.~Feretti, G.~Giovannini, F.~Govoni, M.~Murgia, G.~Taylor et~al., \emph{Revealing the magnetic field in a distant galaxy cluster: discovery of the complex radio emission from macs j0717. 5+ 3745}, {\emph{Astronomy \& Astrophysics} {\bfseries 503} (2009) 707}.

\bibitem{eckert2016xxl}
D.~Eckert, S.~Ettori, J.~Coupon, F.~Gastaldello, M.~Pierre, J.-B.~Melin et~al., \emph{The xxl survey-xiii. baryon content of the bright cluster sample}, {\emph{Astronomy \& Astrophysics} {\bfseries 592} (2016) A12}.

\bibitem{1986rpa..book.....R}
G.B.~{Rybicki} and A.P.~{Lightman}, \emph{{Radiative Processes in Astrophysics}} (1986).

\bibitem{prestage1988cluster}
R.M.~Prestage and J.A.~Peacock, \emph{The cluster environments of powerful radio galaxies}, {\emph{Monthly Notices of the Royal Astronomical Society} {\bfseries 230} (1988) 131}.

\bibitem{kaastra2004spatially}
J.~Kaastra, T.~Tamura, J.~Peterson, J.~Bleeker, C.~Ferrigno, S.~Kahn et~al., \emph{Spatially resolved x-ray spectroscopy of cooling clusters of galaxies}, {\emph{Astronomy \& Astrophysics} {\bfseries 413} (2004) 415}.

\bibitem{carilli2002cluster}
C.~Carilli and G.~Taylor, \emph{Cluster magnetic fields}, {\emph{Annual Review of Astronomy and Astrophysics} {\bfseries 40} (2002) 319}.

\bibitem{GOVONI_2004}
F.~GOVONI and L.~FERETTI, \emph{Magnetic fields in clusters of galaxies}, \href{https://doi.org/10.1142/s0218271804005080}{\emph{International Journal of Modern Physics D} {\bfseries 13} (2004) 1549–1594}.

\bibitem{1972CoASP...4..173S}
R.A.~{Sunyaev} and Y.B.~{Zeldovich}, \emph{{The Observations of Relic Radiation as a Test of the Nature of X-Ray Radiation from the Clusters of Galaxies}}, {\emph{Comments on Astrophysics and Space Physics} {\bfseries 4} (1972) 173}.

\bibitem{sarazin1986x}
C.L.~Sarazin, \emph{X-ray emission from clusters of galaxies}, {\emph{Reviews of Modern Physics} {\bfseries 58} (1986) 1}.

\bibitem{Birkinshaw_1999}
M.~Birkinshaw, \emph{The sunyaev–zel’dovich effect}, \href{https://doi.org/10.1016/s0370-1573(98)00080-5}{\emph{Physics Reports} {\bfseries 310} (1999) 97–195}.

\bibitem{2014ApJS..210....9B}
M.~{Bilicki}, T.H.~{Jarrett}, J.A.~{Peacock}, M.E.~{Cluver} and L.~{Steward}, \emph{{Two Micron All Sky Survey Photometric Redshift Catalog: A Comprehensive Three-dimensional Census of the Whole Sky}}, \href{https://doi.org/10.1088/0067-0049/210/1/9}{\emph{\apjs} {\bfseries 210} (2014) 9} [\href{https://arxiv.org/abs/1311.5246}{{\ttfamily 1311.5246}}].

\bibitem{zehavi2011galaxy}
I.~Zehavi, Z.~Zheng, D.H.~Weinberg, M.R.~Blanton, N.A.~Bahcall, A.A.~Berlind et~al., \emph{Galaxy clustering in the completed sdss redshift survey: the dependence on color and luminosity}, {\emph{The Astrophysical Journal} {\bfseries 736} (2011) 59}.

\bibitem{castagne2012deep}
D.~Castagn{\'e}, G.~Soucail, E.~Pointecouteau, A.~Cappi, S.~Maurogordato, C.~Benoist et~al., \emph{Deep optical observations of the massive galaxy cluster abell 1413}, {\emph{Astronomy \& Astrophysics} {\bfseries 548} (2012) A18}.

\bibitem{abdalla2011comparison}
F.B.~Abdalla, M.~Banerji, O.~Lahav and V.~Rashkov, \emph{A comparison of six photometric redshift methods applied to 1.5 million luminous red galaxies}, {\emph{Monthly Notices of the Royal Astronomical Society} {\bfseries 417} (2011) 1891}.

\bibitem{zaznobin2021spectroscopic}
I.~Zaznobin, R.~Burenin, I.~Bikmaev, I.~Khamitov, G.~Khorunzhev, A.~Lyapin et~al., \emph{Spectroscopic redshift measurements for galaxy clusters from the planck survey and observations of these clusters in the srg/erosita survey}, {\emph{Astronomy Letters} {\bfseries 47} (2021) 61}.

\bibitem{Fixsen_1996}
D.J.~Fixsen, E.S.~Cheng, J.M.~Gales, J.C.~Mather, R.A.~Shafer and E.L.~Wright, \emph{The cosmic microwave background spectrum from the fullcobefiras data set}, \href{https://doi.org/10.1086/178173}{\emph{The Astrophysical Journal} {\bfseries 473} (1996) 576–587}.

\bibitem{fixsen1996cosmic}
D.J.~Fixsen, E.~Cheng, J.~Gales, J.C.~Mather, R.~Shafer and E.~Wright, \emph{The cosmic microwave background spectrum from the full cobe* firas data set}, {\emph{The Astrophysical Journal} {\bfseries 473} (1996) 576}.

\bibitem{hanson2009estimators}
D.~Hanson and A.~Lewis, \emph{Estimators for cmb statistical anisotropy}, {\emph{Physical Review D—Particles, Fields, Gravitation, and Cosmology} {\bfseries 80} (2009) 063004}.

\bibitem{Fixsen_2009}
D.J.~Fixsen, \emph{The temperature of the cosmic microwave background}, \href{https://doi.org/10.1088/0004-637x/707/2/916}{\emph{The Astrophysical Journal} {\bfseries 707} (2009) 916–920}.

\bibitem{hu2002cosmic}
W.~Hu and S.~Dodelson, \emph{Cosmic microwave background anisotropies}, {\emph{Annual Review of Astronomy and Astrophysics} {\bfseries 40} (2002) 171}.

\bibitem{adam2016planck}
R.~Adam, N.~Aghanim, M.~Ashdown, J.~Aumont, C.~Baccigalupi, M.~Ballardini et~al., \emph{Planck intermediate results-xlvii. planck constraints on reionization history}, {\emph{Astronomy \& Astrophysics} {\bfseries 596} (2016) A108}.

\bibitem{Smith_2007}
K.M.~Smith, O.~Zahn and O.~Doré, \emph{Detection of gravitational lensing in the cosmic microwave background}, \href{https://doi.org/10.1103/physrevd.76.043510}{\emph{Physical Review D} {\bfseries 76} (2007) }.

\bibitem{2014PTEP.2014fB107T}
H.~{Tashiro}, \emph{{CMB spectral distortions and energy release in the early universe}}, \href{https://doi.org/10.1093/ptep/ptu066}{\emph{Progress of Theoretical and Experimental Physics} {\bfseries 2014} (2014) 06B107}.

\bibitem{erler2018planck}
J.~Erler, K.~Basu, J.~Chluba and F.~Bertoldi, \emph{Planck's view on the spectrum of the sunyaev--zeldovich effect}, {\emph{Monthly Notices of the Royal Astronomical Society} {\bfseries 476} (2018) 3360}.

\bibitem{dine1983not}
M.~Dine and W.~Fischler, \emph{The not-so-harmless axion}, {\emph{Physics Letters B} {\bfseries 120} (1983) 137}.

\bibitem{abbott1983cosmological}
L.F.~Abbott and P.~Sikivie, \emph{A cosmological bound on the invisible axion}, {\emph{Physics Letters B} {\bfseries 120} (1983) 133}.

\bibitem{preskill1983cosmology}
J.~Preskill, M.B.~Wise and F.~Wilczek, \emph{Cosmology of the invisible axion}, {\emph{Physics Letters B} {\bfseries 120} (1983) 127}.

\bibitem{Ghosh:2022rta}
A.~Ghosh, P.~Konar and R.~Roshan, \emph{{Top-philic dark matter in a hybrid KSVZ axion framework}}, \href{https://doi.org/10.1007/JHEP12(2022)167}{\emph{JHEP} {\bfseries 12} (2022) 167} [\href{https://arxiv.org/abs/2207.00487}{{\ttfamily 2207.00487}}].

\bibitem{1992SvJNP..55.1063B}
Z.G.~{Berezhiani}, A.S.~{Sakharov} and M.Y.~{Khlopov}, \emph{{Primordial background of cosmological axions.}}, {\emph{Soviet Journal of Nuclear Physics} {\bfseries 55} (1992) 1063}.

\bibitem{khlopov1999nonlinear}
M.Y.~Khlopov, A.~Sakharov and D.~Sokoloff, \emph{The nonlinear modulation of the density distribution in standard axionic cdm and its cosmological impact}, {\emph{Nuclear Physics B-Proceedings Supplements} {\bfseries 72} (1999) 105}.

\bibitem{sakharov1994nonhomogeneity}
A.~Sakharov and M.Y.~Khlopov, \emph{The nonhomogeneity problem for the primordial axion field}, {\emph{Physics of Atomic Nuclei} {\bfseries 57} (1994) 485}.

\bibitem{sakharov1996large}
A.~Sakharov, D.~Sokoloff and M.Y.~Khlopov, \emph{Large-scale modulation of the distribution of coherent oscillations of a primordial axion field in the universe}, {\emph{Physics of Atomic Nuclei} {\bfseries 59} (1996) }.

\bibitem{Mukherjee_2020}
S.~Mukherjee, D.N.~Spergel, R.~Khatri and B.D.~Wandelt, \emph{A new probe of axion-like particles: Cmb polarization distortions due to cluster magnetic fields}, \href{https://doi.org/10.1088/1475-7516/2020/02/032}{\emph{Journal of Cosmology and Astroparticle Physics} {\bfseries 2020} (2020) 032–032}.

\bibitem{Mehta:2024wfo}
H.~Mehta and S.~Mukherjee, \emph{{A power spectrum approach to the search for axion-like particles from resolved galaxy clusters using CMB as a backlight}}, \href{https://doi.org/10.1088/1475-7516/2024/09/037}{\emph{JCAP} {\bfseries 09} (2024) 037} [\href{https://arxiv.org/abs/2405.08878}{{\ttfamily 2405.08878}}].

\bibitem{Mehta:2024pdz}
H.~Mehta and S.~Mukherjee, \emph{{A diffused background from axion-like particles in the microwave sky}}, \href{https://doi.org/10.1088/1475-7516/2024/07/084}{\emph{JCAP} {\bfseries 07} (2024) 084} [\href{https://arxiv.org/abs/2405.08879}{{\ttfamily 2405.08879}}].

\bibitem{Carosi:2013rla}
G.~Carosi, A.~Friedland, M.~Giannotti, M.J.~Pivovaroff, J.~Ruz and J.K.~Vogel, \emph{{Probing the axion-photon coupling: phenomenological and experimental perspectives. A snowmass white paper}},  in \emph{{Snowmass 2013}: {Snowmass on the Mississippi}}, 9, 2013 [\href{https://arxiv.org/abs/1309.7035}{{\ttfamily 1309.7035}}].

\bibitem{Ghosh:2023xhs}
A.~Ghosh and P.~Konar, \emph{{Precision prediction at the LHC of a democratic up-family philic KSVZ axion model}},  \href{https://arxiv.org/abs/2305.08662}{{\ttfamily 2305.08662}}.

\bibitem{Mukherjee_2018}
S.~Mukherjee, R.~Khatri and B.D.~Wandelt, \emph{Polarized anisotropic spectral distortions of the cmb: galactic and extragalactic constraints on photon-axion conversion}, \href{https://doi.org/10.1088/1475-7516/2018/04/045}{\emph{Journal of Cosmology and Astroparticle Physics} {\bfseries 2018} (2018) 045–045}.

\bibitem{Mukherjee_2019}
S.~Mukherjee, R.~Khatri and B.D.~Wandelt, \emph{Constraints on non-resonant photon-axion conversion from the planck satellite data}, \href{https://doi.org/10.1088/1475-7516/2019/06/031}{\emph{Journal of Cosmology and Astroparticle Physics} {\bfseries 2019} (2019) 031–031}.

\bibitem{osti_22525054}
M.~Schlederer and G.~Sigl, \emph{Constraining alp-photon coupling using galaxy clusters}, \href{https://doi.org/10.1088/1475-7516/2016/01/038}{\emph{Journal of Cosmology and Astroparticle Physics} {\bfseries 2016} (2016) }.

\bibitem{salvato2018finding}
M.~Salvato, J.~Buchner, T.~Budav{\'a}ri, T.~Dwelly, A.~Merloni, M.~Brusa et~al., \emph{Finding counterparts for all-sky x-ray surveys with nway: a bayesian algorithm for cross-matching multiple catalogues}, {\emph{Monthly Notices of the Royal Astronomical Society} {\bfseries 473} (2018) 4937}.

\bibitem{2016}
P.A.R.~Ade, N.~Aghanim, M.~Arnaud, M.~Ashdown and J.e.a.~Aumont, \emph{Planck2015 results: Xiii. cosmological parameters}, \href{https://doi.org/10.1051/0004-6361/201525830}{\emph{Astronomy \&; Astrophysics} {\bfseries 594} (2016) A13}.

\bibitem{mohr1999properties}
J.J.~Mohr, B.~Mathiesen and A.E.~Evrard, \emph{Properties of the intracluster medium in an ensemble of nearby galaxy clusters}, {\emph{The Astrophysical Journal} {\bfseries 517} (1999) 627}.

\bibitem{Ade_2019}
P.~Ade, J.~Aguirre and Z.e.a.~Ahmed, \emph{The simons observatory: science goals and forecasts}, \href{https://doi.org/10.1088/1475-7516/2019/02/056}{\emph{Journal of Cosmology and Astroparticle Physics} {\bfseries 2019} (2019) 056–056}.

\bibitem{abazajian2016cmbs4}
K.N.~Abazajian, P.~Adshead, Z.~Ahmed and S.W.A.~et~al., \emph{Cmb-s4 science book, first edition},  2016.

\bibitem{Carilli_2004}
C.~Carilli and S.~Rawlings, \emph{Motivation, key science projects, standards and assumptions}, \href{https://doi.org/10.1016/j.newar.2004.09.001}{\emph{New Astronomy Reviews} {\bfseries 48} (2004) 979–984}.

\bibitem{braun2019anticipated}
R.~Braun, A.~Bonaldi, T.~Bourke, E.~Keane and J.~Wagg, \emph{Anticipated performance of the square kilometre array--phase 1 (ska1)}, {\emph{arXiv preprint arXiv:1912.12699} (2019) }.

\bibitem{brown2004alma}
R.L.~Brown, W.~Wild and C.~Cunningham, \emph{Alma--the atacama large millimeter array}, {\emph{Advances in Space Research} {\bfseries 34} (2004) 555}.

\bibitem{predehl2021erosita}
P.~Predehl, R.~Andritschke, V.~Arefiev, V.~Babyshkin, O.~Batanov, W.~Becker et~al., \emph{The erosita x-ray telescope on srg}, {\emph{Astronomy \& Astrophysics} {\bfseries 647} (2021) A1}.

\bibitem{gardner2006james}
J.P.~Gardner, J.C.~Mather, M.~Clampin, R.~Doyon, M.A.~Greenhouse, H.B.~Hammel et~al., \emph{The james webb space telescope}, {\emph{Space Science Reviews} {\bfseries 123} (2006) 485}.

\bibitem{sehgal2019cmbhd}
N.~Sehgal, S.~Aiola, Y.~Akrami and K.B.~et~al., \emph{Cmb-hd: An ultra-deep, high-resolution millimeter-wave survey over half the sky},  \href{https://arxiv.org/abs/1906.10134}{{\ttfamily 1906.10134}}.

\bibitem{condon2016essential}
J.J.~Condon and S.M.~Ransom, \emph{Essential radio astronomy}, vol.~2, Princeton University Press (2016).

\bibitem{clarke2001new}
T.E.~Clarke, P.P.~Kronberg and H.~B{\"o}hringer, \emph{A new radio-x-ray probe of galaxy cluster magnetic fields}, {\emph{The Astrophysical Journal} {\bfseries 547} (2001) L111}.

\bibitem{Ferrari_2008}
C.~Ferrari, F.~Govoni, S.~Schindler, A.M.~Bykov and Y.~Rephaeli, \emph{Observations of extended radio emission in clusters}, \href{https://doi.org/10.1007/s11214-008-9311-x}{\emph{Space Science Reviews} {\bfseries 134} (2008) 93–118}.

\bibitem{ghatak2009optics}
A.~Ghatak, \emph{Optics}, Tata McGraw-Hill Publishing Company Limited (2009).

\bibitem{murgia2004magnetic}
M.~Murgia, F.~Govoni, L.~Feretti, G.~Giovannini, D.~Dallacasa, R.~Fanti et~al., \emph{Magnetic fields and faraday rotation in clusters of galaxies}, {\emph{Astronomy \& Astrophysics} {\bfseries 424} (2004) 429}.

\bibitem{bohringer2016cosmic}
H.~B{\"o}hringer, G.~Chon and P.P.~Kronberg, \emph{The cosmic large-scale structure in x-rays (classix) cluster survey-i. probing galaxy cluster magnetic fields with line of sight rotation measures}, {\emph{Astronomy \& Astrophysics} {\bfseries 596} (2016) A22}.

\bibitem{eilek2002magnetic}
J.A.~Eilek and F.N.~Owen, \emph{Magnetic fields in cluster cores: Faraday rotation in a400 and a2634}, {\emph{The Astrophysical Journal} {\bfseries 567} (2002) 202}.

\bibitem{bonafede2010galaxy}
A.~Bonafede, L.~Feretti, M.~Murgia, F.~Govoni, G.~Giovannini and V.~Vacca, \emph{Galaxy cluster magnetic fields from radio polarized emission}, {\emph{arXiv preprint arXiv:1009.1233} (2010) }.

\bibitem{loi2017magnetic}
F.~Loi, F.~Govoni, M.~Murgia, V.~Vacca, H.~Li, L.~Feretti et~al., \emph{Magnetic fields in galaxy clusters in the ska era},  in \emph{Journal of Physics: Conference Series}, vol.~841, p.~012005, IOP Publishing, 2017.

\bibitem{bulbul2024srgerosita}
E.~Bulbul, A.~Liu, M.~Kluge, X.~Zhang, J.S.~Sanders, Y.E.~Bahar et~al., \emph{The srg/erosita all-sky survey: The first catalog of galaxy clusters and groups in the western galactic hemisphere},  \href{https://arxiv.org/abs/2402.08452}{{\ttfamily 2402.08452}}.

\bibitem{merloni2012erosita}
A.~Merloni, P.~Predehl, W.~Becker, H.~Böhringer, T.~Boller and H.B.~et~al., \emph{erosita science book: Mapping the structure of the energetic universe},  \href{https://arxiv.org/abs/1209.3114}{{\ttfamily 1209.3114}}.

\bibitem{pratt2016hot}
G.~Pratt, E.~Pointecouteau, M.~Arnaud and R.~Van~der Burg, \emph{The hot gas content of fossil galaxy clusters}, {\emph{Astronomy \& Astrophysics} {\bfseries 590} (2016) L1}.

\bibitem{mcdonald2013growth}
M.~McDonald, B.~Benson, A.~Vikhlinin, B.~Stalder, L.~Bleem, T.~De~Haan et~al., \emph{The growth of cool cores and evolution of cooling properties in a sample of 83 galaxy clusters at 0.3< z< 1.2 selected from the spt-sz survey}, {\emph{The Astrophysical Journal} {\bfseries 774} (2013) 23}.

\bibitem{mcdonald2017remarkable}
M.~McDonald, S.~Allen, M.~Bayliss, B.~Benson, L.~Bleem, M.~Brodwin et~al., \emph{The remarkable similarity of massive galaxy clusters from z~ 0 to z~ 1.9}, {\emph{The Astrophysical Journal} {\bfseries 843} (2017) 28}.

\bibitem{felten1966x}
J.~Felten, R.~Gould, W.~Stein and N.~Woolf, \emph{X-rays from the coma cluster of galaxies}, {\emph{Astrophysical Journal, vol. 146, p. 955-958} {\bfseries 146} (1966) 955}.

\bibitem{cavaliere1978distribution}
A.~Cavaliere and R.~Fusco-Femiano, \emph{The distribution of hot gas in clusters of galaxies}, {\emph{Astronomy and Astrophysics, Vol. 70, p. 677 (1978)} {\bfseries 70} (1978) 677}.

\bibitem{Vikhlinin_2005}
A.~Vikhlinin, M.~Markevitch, S.S.~Murray, C.~Jones, W.~Forman and L.~Van~Speybroeck, \emph{Chandratemperature profiles for a sample of nearby relaxed galaxy clusters}, \href{https://doi.org/10.1086/431142}{\emph{The Astrophysical Journal} {\bfseries 628} (2005) 655–672}.

\bibitem{Vikhlinin_2006}
A.~Vikhlinin, A.~Kravtsov, W.~Forman, C.~Jones, M.~Markevitch, S.S.~Murray et~al., \emph{Chandrasample of nearby relaxed galaxy clusters: Mass, gas fraction, and mass‐temperature relation}, \href{https://doi.org/10.1086/500288}{\emph{The Astrophysical Journal} {\bfseries 640} (2006) 691–709}.

\bibitem{fadley2014some}
C.S.~Fadley and S.~Nem{\v{s}}{\'a}k, \emph{Some future perspectives in soft-and hard-x-ray photoemission}, {\emph{Journal of Electron Spectroscopy and Related Phenomena} {\bfseries 195} (2014) 409}.

\bibitem{Staniszewski_2009}
Z.~Staniszewski, P.A.R.~Ade, K.A.~Aird, B.A.~Benson, L.E.~Bleem, J.E.~Carlstrom et~al., \emph{Galaxy clusters discovered with a sunyaev-zel’dovich effect survey}, \href{https://doi.org/10.1088/0004-637x/701/1/32}{\emph{The Astrophysical Journal} {\bfseries 701} (2009) 32–41}.

\bibitem{komatsu1999submillimeter}
E.~Komatsu, T.~Kitayama, Y.~Suto, M.~Hattori, R.~Kawabe, H.~Matsuo et~al., \emph{Submillimeter detection of the sunyaev-zeldovich effect toward the most luminous x-ray cluster at z= 0.45}, {\emph{The Astrophysical Journal} {\bfseries 516} (1999) L1}.

\bibitem{sunyaev1980microwave}
R.~Sunyaev and I.B.~Zeldovich, \emph{Microwave background radiation as a probe of the contemporary structure and history of the universe}, {\emph{In: Annual review of astronomy and astrophysics. Volume 18.(A81-20334 07-90) Palo Alto, Calif., Annual Reviews, Inc., 1980, p. 537-560.} {\bfseries 18} (1980) 537}.

\bibitem{battaglia2016tau}
N.~Battaglia, \emph{The tau of galaxy clusters}, {\emph{Journal of Cosmology and Astroparticle Physics} {\bfseries 2016} (2016) 058}.

\bibitem{carlstrom2002cosmology}
J.E.~Carlstrom, G.P.~Holder and E.D.~Reese, \emph{Cosmology with the sunyaev-zel’dovich effect}, {\emph{Annual Review of Astronomy and Astrophysics} {\bfseries 40} (2002) 643}.

\bibitem{Komatsu_1999}
E.~Komatsu and T.~Kitayama, \emph{Sunyaev-zeldovich fluctuations from spatial correlations between clusters of galaxies}, \href{https://doi.org/10.1086/312364}{\emph{The Astrophysical Journal} {\bfseries 526} (1999) L1–L4}.

\bibitem{adam2017mapping}
R.~Adam, M.~Arnaud, I.~Bartalucci, P.~Ade, P.~Andr{\'e}, A.~Beelen et~al., \emph{Mapping the hot gas temperature in galaxy clusters using x-ray and sunyaev-zel’dovich imaging}, {\emph{Astronomy \& Astrophysics} {\bfseries 606} (2017) A64}.

\bibitem{shitanishi2018thermodynamic}
J.A.~Shitanishi, E.~Pierpaoli, J.~Sayers, S.R.~Golwala, S.~Ameglio, A.B.~Mantz et~al., \emph{Thermodynamic profiles of galaxy clusters from a joint x-ray/sz analysis}, {\emph{Monthly Notices of the Royal Astronomical Society} {\bfseries 481} (2018) 749}.

\bibitem{bhatiani2022optical}
S.~Bhatiani, X.~Dai, R.D.~Griffin, J.M.~Nugent, C.S.~Kochanek and J.N.~Bregman, \emph{Optical confirmation of x-ray-selected galaxy clusters from the swift agn and cluster survey with mdm and pan-starrs data. iii}, {\emph{The Astrophysical Journal Supplement Series} {\bfseries 259} (2022) 9}.

\bibitem{csabai2003application}
I.~Csabai, T.~Budavari, A.J.~Connolly, A.S.~Szalay, Z.~Gy{\H{o}}ry, N.~Benitez et~al., \emph{The application of photometric redshifts to the sdss early data release}, {\emph{The Astronomical Journal} {\bfseries 125} (2003) 580}.

\bibitem{bilicki2016wise}
M.~Bilicki, J.A.~Peacock, T.H.~Jarrett, M.E.~Cluver, N.~Maddox, M.J.~Brown et~al., \emph{Wise$\times$ supercosmos photometric redshift catalog: 20 million galaxies over 3$\pi$ steradians}, {\emph{The Astrophysical Journal Supplement Series} {\bfseries 225} (2016) 5}.

\bibitem{gilbank2004exploring}
D.G.~Gilbank, R.G.~Bower, F.~Castander and B.~Ziegler, \emph{Exploring the selection of galaxy clusters and groups: an optical survey for x-ray dark clusters}, {\emph{Monthly Notices of the Royal Astronomical Society} {\bfseries 348} (2004) 551}.

\bibitem{yee2001optical}
H.~Yee and M.~Gladders, \emph{Optical surveys for galaxy clusters}, {\emph{arXiv preprint astro-ph/0111431} (2001) }.

\bibitem{wen2018catalogue}
Z.~Wen, J.~Han and F.~Yang, \emph{A catalogue of clusters of galaxies identified from all sky surveys of 2mass, wise, and supercosmos}, {\emph{Monthly Notices of the Royal Astronomical Society} {\bfseries 475} (2018) 343}.

\bibitem{girardi1998optical}
M.~Girardi, G.~Giuricin, F.~Mardirossian, M.~Mezzetti and W.~Boschin, \emph{Optical mass estimates of galaxy clusters}, {\emph{The Astrophysical Journal} {\bfseries 505} (1998) 74}.

\bibitem{berezhiani1991cosmology}
Z.~Berezhiani and M.Y.~Khlopov, \emph{Cosmology of spontaneously broken gauge family symmetry with axion solution of strong cp-problem}, {\emph{Zeitschrift f{\"u}r Physik C Particles and Fields} {\bfseries 49} (1991) 73}.

\bibitem{chadha2022axion}
F.~Chadha-Day, J.~Ellis and D.J.~Marsh, \emph{Axion dark matter: What is it and why now?}, {\emph{Science advances} {\bfseries 8} (2022) eabj3618}.

\bibitem{Raffelt:1996wa}
G.G.~Raffelt, \emph{{Stars as laboratories for fundamental physics}: {The astrophysics of neutrinos, axions, and other weakly interacting particles}} (5, 1996).

\bibitem{Choi_2021}
K.~Choi, S.H.~Im and C.S.~Shin, \emph{Recent progress in the physics of axions and axion-like particles}, \href{https://doi.org/10.1146/annurev-nucl-120720-031147}{\emph{Annual Review of Nuclear and Particle Science} {\bfseries 71} (2021) 225–252}.

\bibitem{smirnov2005msw}
A.Y.~Smirnov, \emph{The msw effect and matter effects in neutrino oscillations}, {\emph{Physica Scripta} {\bfseries 2005} (2005) 57}.

\bibitem{langacker1987implications}
P.~Langacker, S.~Petcov, G.~Steigman and S.~Toshev, \emph{Implications of the mikheyev-smirnov-wolfenstein (msw) mechanism of amplification of neutrino oscillations in matter}, {\emph{Nuclear Physics B} {\bfseries 282} (1987) 589}.

\bibitem{forero2014neutrino}
D.~Forero, M.~Tortola and J.~Valle, \emph{Neutrino oscillations refitted}, {\emph{Physical Review D} {\bfseries 90} (2014) 093006}.

\bibitem{duan2010collective}
H.~Duan, G.M.~Fuller and Y.-Z.~Qian, \emph{Collective neutrino oscillations}, {\emph{Annual Review of Nuclear and Particle Science} {\bfseries 60} (2010) 569}.

\bibitem{wolfenstein2018neutrino}
L.~Wolfenstein, \emph{Neutrino oscillations in matter},  in \emph{Solar neutrinos}, pp.~294--299, CRC Press (2018).

\bibitem{kuo1989neutrino}
T.-K.~Kuo and J.~Pantaleone, \emph{Neutrino oscillations in matter}, {\emph{Reviews of Modern Physics} {\bfseries 61} (1989) 937}.

\bibitem{bilenky1987massive}
S.M.~Bilenky and S.~Petcov, \emph{Massive neutrinos and neutrino oscillations}, {\emph{Reviews of Modern Physics} {\bfseries 59} (1987) 671}.

\bibitem{bilenky1999phenomenology}
S.M.~Bilenky, C.~Giunti and W.~Grimus, \emph{Phenomenology of neutrino oscillations}, {\emph{Progress in Particle and Nuclear Physics} {\bfseries 43} (1999) 1}.

\bibitem{marsh2016axion}
D.J.~Marsh, \emph{Axion cosmology}, {\emph{Physics Reports} {\bfseries 643} (2016) 1}.

\bibitem{perna2012signatures}
R.~Perna, W.C.~Ho, L.~Verde, M.~Van~Adelsberg and R.~Jimenez, \emph{Signatures of photon--axion conversion in the thermal spectra and polarization of neutron stars}, {\emph{The Astrophysical Journal} {\bfseries 748} (2012) 116}.

\bibitem{bondarenko2023neutron}
K.~Bondarenko, A.~Boyarsky, J.~Pradler and A.~Sokolenko, \emph{Neutron stars as photon double-lenses: constraining resonant conversion into alps}, {\emph{Physics Letters B} {\bfseries 846} (2023) 138238}.

\bibitem{lella2023protoneutron}
A.~Lella, P.~Carenza, G.~Lucente, M.~Giannotti and A.~Mirizzi, \emph{Protoneutron stars as cosmic factories for massive axionlike particles}, {\emph{Physical Review D} {\bfseries 107} (2023) 103017}.

\bibitem{yu2023searching}
Q.~Yu and D.~Horns, \emph{Searching for photon-alps mixing effects in agn gamma-ray energy spectra}, {\emph{Journal of Cosmology and Astroparticle Physics} {\bfseries 2023} (2023) 029}.

\bibitem{Mirizzi_2009}
A.~Mirizzi, J.~Redondo and G.~Sigl, \emph{Constraining resonant photon-axion conversions in the early universe}, \href{https://doi.org/10.1088/1475-7516/2009/08/001}{\emph{Journal of Cosmology and Astroparticle Physics} {\bfseries 2009} (2009) 001–001}.

\bibitem{2017}
\emph{New cast limit on the axion–photon interaction}, \href{https://doi.org/10.1038/nphys4109}{\emph{Nature Physics} {\bfseries 13} (2017) 584–590}.

\bibitem{dvorkin2022physicslightrelics}
C.~Dvorkin, J.~Meyers, P.~Adshead, M.~Amin, C.A.~Argüelles, T.~Brinckmann et~al., \emph{The physics of light relics},  2022.

\bibitem{https://doi.org/10.17863/cam.30368}
B.~Wallisch, \emph{Cosmological Probes of Light Relics}, Ph.D. thesis, 2018.
\newblock 10.17863/CAM.30368.

\bibitem{green2019messengersearlyuniversecosmic}
D.~Green, M.A.~Amin, J.~Meyers, B.~Wallisch, K.N.~Abazajian, M.~Abidi et~al., \emph{Messengers from the early universe: Cosmic neutrinos and other light relics},  2019.

\bibitem{Hu_2002}
W.~Hu and S.~Dodelson, \emph{Cosmic microwave background anisotropies}, \href{https://doi.org/10.1146/annurev.astro.40.060401.093926}{\emph{Annual Review of Astronomy and Astrophysics} {\bfseries 40} (2002) 171–216}.

\bibitem{white1996sachs}
M.~White and W.~Hu, \emph{The sachs-wolfe effect}, {\emph{arXiv preprint astro-ph/9609105} (1996) }.

\bibitem{mondino2024axioninducedpatchyscreeningcosmic}
C.~Mondino, D.~Pîrvu, J.~Huang and M.C.~Johnson, \emph{Axion-induced patchy screening of the cosmic microwave background},  \href{https://arxiv.org/abs/2405.08059}{{\ttfamily 2405.08059}}.

\bibitem{Foreman_Mackey_2013}
D.~Foreman-Mackey, D.W.~Hogg, D.~Lang and J.~Goodman, \emph{<tt>emcee</tt>: The mcmc hammer}, \href{https://doi.org/10.1086/670067}{\emph{Publications of the Astronomical Society of the Pacific} {\bfseries 125} (2013) 306–312}.

\bibitem{osinga2024probing}
E.~Osinga, R.J.~van Weeren, L.~Rudnick, F.~Andrade-Santos, A.~Bonafede, T.~Clarke et~al., \emph{Probing cluster magnetism with embedded and background radio sources in planck clusters}, {\emph{arXiv preprint arXiv:2408.07178} (2024) }.

\bibitem{heald2020magnetism}
G.~Heald, S.A.~Mao, V.~Vacca, T.~Akahori, A.~Damas-Segovia, B.~Gaensler et~al., \emph{Magnetism science with the square kilometre array}, {\emph{Galaxies} {\bfseries 8} (2020) 53}.

\bibitem{bartalucci2017recovering}
I.~Bartalucci, M.~Arnaud, G.~Pratt, A.~Vikhlinin, E.~Pointecouteau, W.~Forman et~al., \emph{Recovering galaxy cluster gas density profiles with xmm-newton and chandra}, {\emph{Astronomy \& Astrophysics} {\bfseries 608} (2017) A88}.

\bibitem{de2002temperature}
S.~De~Grandi and S.~Molendi, \emph{Temperature profiles of nearby clusters of galaxies}, {\emph{The Astrophysical Journal} {\bfseries 567} (2002) 163}.

\bibitem{kafer2019toward}
F.~K{\"a}fer, A.~Finoguenov, D.~Eckert, J.S.~Sanders, T.H.~Reiprich and K.~Nandra, \emph{Toward a characterization of x-ray galaxy clusters for cosmology}, {\emph{Astronomy \& Astrophysics} {\bfseries 628} (2019) A43}.

\bibitem{Zonca2019}
A.~Zonca, L.~Singer, D.~Lenz, M.~Reinecke, C.~Rosset, E.~Hivon et~al., \emph{healpy: equal area pixelization and spherical harmonics transforms for data on the sphere in python}, \href{https://doi.org/10.21105/joss.01298}{\emph{Journal of Open Source Software} {\bfseries 4} (2019) 1298}.

\bibitem{Thorne_2017}
B.~Thorne, J.~Dunkley, D.~Alonso and S.~Næss, \emph{The python sky model: software for simulating the galactic microwave sky}, \href{https://doi.org/10.1093/mnras/stx949}{\emph{Monthly Notices of the Royal Astronomical Society} {\bfseries 469} (2017) 2821–2833}.

\bibitem{Sazonov_1999}
S.Y.~Sazonov and R.A.~Sunyaev, \emph{Microwave polarization in the direction of galaxy clusters induced by the cmb quadrupole anisotropy}, \href{https://doi.org/10.1046/j.1365-8711.1999.02981.x}{\emph{Monthly Notices of the Royal Astronomical Society} {\bfseries 310} (1999) 765–772}.

\bibitem{shimon2009power}
M.~Shimon, Y.~Rephaeli, S.~Sadeh and B.~Keating, \emph{Power spectra of cmb polarization by scattering in clusters}, {\emph{Monthly Notices of the Royal Astronomical Society} {\bfseries 399} (2009) 2088}.

\bibitem{Louis_2017}
T.~Louis, E.F.~Bunn, B.~Wandelt and J.~Silk, \emph{Measuring polarized emission in clusters in the cmb s4 era}, \href{https://doi.org/10.1103/physrevd.96.123509}{\emph{Physical Review D} {\bfseries 96} (2017) }.

\bibitem{Hall_2014}
A.~Hall and A.~Challinor, \emph{Detecting the polarization induced by scattering of the microwave background quadrupole in galaxy clusters}, \href{https://doi.org/10.1103/physrevd.90.063518}{\emph{Physical Review D} {\bfseries 90} (2014) }.

\bibitem{aghanim2008secondary}
N.~Aghanim, S.~Majumdar and J.~Silk, \emph{Secondary anisotropies of the cmb}, {\emph{Reports on Progress in Physics} {\bfseries 71} (2008) 066902}.

\bibitem{yasini2016kinetic}
S.~Yasini and E.~Pierpaoli, \emph{Kinetic sunyaev-zeldovich effect in an anisotropic cmb model: Measuring low multipoles of the cmb at higher redshifts using intensity and polarization spectral distortions}, {\emph{Physical Review D} {\bfseries 94} (2016) 023513}.

\bibitem{challinor2000thermal}
A.~Challinor, M.~Ford and A.~Lasenby, \emph{Thermal and kinematic corrections to the microwave background polarization induced by galaxy clusters along the line of sight}, {\emph{Monthly Notices of the Royal Astronomical Society} {\bfseries 312} (2000) 159}.

\bibitem{ilc2008internal}
R.~Vio and P.~Andreani, \emph{"internal linear combination" method for the separation of cmb from galactic foregrounds in the harmonic domain},  \href{https://arxiv.org/abs/0811.4277}{{\ttfamily 0811.4277}}.

\bibitem{Eriksen_2004}
H.K.~Eriksen, A.J.~Banday, K.M.~Gorski and P.B.~Lilje, \emph{On foreground removal from thewilkinson microwave anisotropy probedata by an internal linear combination method: Limitations and implications}, \href{https://doi.org/10.1086/422807}{\emph{The Astrophysical Journal} {\bfseries 612} }.

\bibitem{Hansen_2006}
F.K.~Hansen, A.J.~Banday, H.K.~Eriksen, K.M.~Gorski and P.B.~Lilje, \emph{Foreground subtraction of cosmic microwave background maps using wi‐fit (wavelet‐based high‐resolution fitting of internal templates)}, \href{https://doi.org/10.1086/506015}{\emph{The Astrophysical Journal} {\bfseries 648} (2006) 784–796}.

\bibitem{2020_planck}
Y.~Akrami, M.~Ashdown, J.~Aumont, C.~Baccigalupi, M.~Ballardini, A.J.~Banday et~al., \emph{Planck2018 results: Vii. isotropy and statistics of the cmb}, \href{https://doi.org/10.1051/0004-6361/201935201}{\emph{Astronomy \&; Astrophysics} {\bfseries 641} (2020) A7}.

\bibitem{Austermann_2012}
J.E.~Austermann, K.A.~Aird, J.A.~Beall, D.~Becker, A.~Bender, B.A.~Benson et~al., \emph{Sptpol: an instrument for cmb polarization measurements with the south pole telescope},  in \emph{Millimeter, Submillimeter, and Far-Infrared Detectors and Instrumentation for Astronomy VI}, W.S.~Holland, ed., SPIE, Sept., 2012, \href{https://doi.org/10.1117/12.927286}{DOI}.

\bibitem{niemack2010actpol}
M.D.~Niemack, P.A.~Ade, J.~Aguirre, F.~Barrientos, J.~Beall, J.~Bond et~al., \emph{Actpol: a polarization-sensitive receiver for the atacama cosmology telescope},  in \emph{Millimeter, Submillimeter, and Far-Infrared Detectors and Instrumentation for Astronomy V}, vol.~7741, pp.~537--557, SPIE, 2010.

\bibitem{thornton2016atacama}
R.~Thornton, P.~Ade, S.~Aiola, F.~Angile, M.~Amiri, J.~Beall et~al., \emph{The atacama cosmology telescope: the polarization-sensitive actpol instrument}, {\emph{The Astrophysical Journal Supplement Series} {\bfseries 227} (2016) 21}.

\bibitem{astropy:2013}
{Astropy Collaboration}, T.P.~{Robitaille}, E.J.~{Tollerud}, P.~{Greenfield}, M.~{Droettboom} and E.e.a.~{Bray}, \emph{{Astropy: A community Python package for astronomy}}, \href{https://doi.org/10.1051/0004-6361/201322068}{\emph{\aap} {\bfseries 558} (2013) A33} [\href{https://arxiv.org/abs/1307.6212}{{\ttfamily 1307.6212}}].

\bibitem{astropy:2022}
{Astropy Collaboration}, A.M.~{Price-Whelan}, P.L.~{Lim} and N.e.a.~{Earl}, \emph{{The Astropy Project: Sustaining and Growing a Community-oriented Open-source Project and the Latest Major Release (v5.0) of the Core Package}}, \href{https://doi.org/10.3847/1538-4357/ac7c74}{\emph{\apj} {\bfseries 935} (2022) 167} [\href{https://arxiv.org/abs/2206.14220}{{\ttfamily 2206.14220}}].

\bibitem{astropy:2018}
{Astropy Collaboration}, A.M.~{Price-Whelan}, B.M.~{Sip{\H{o}}cz}, H.M.~{G{\"u}nther}, P.L.~{Lim} and S.M.e.a.~{Crawford}, \emph{{The Astropy Project: Building an Open-science Project and Status of the v2.0 Core Package}}, \href{https://doi.org/10.3847/1538-3881/aabc4f}{\emph{\aj} {\bfseries 156} (2018) 123} [\href{https://arxiv.org/abs/1801.02634}{{\ttfamily 1801.02634}}].

\bibitem{harris2020array}
C.R.~Harris, K.J.~Millman, S.J.~van~der Walt, R.~Gommers, P.~Virtanen and D.C.~et~al., \emph{Array programming with {NumPy}}, \href{https://doi.org/10.1038/s41586-020-2649-2}{\emph{Nature} {\bfseries 585} (2020) 357}.

\bibitem{2011ascl.soft02026L}
A.~{Lewis} and A.~{Challinor}, ``{CAMB: Code for Anisotropies in the Microwave Background}.'' Astrophysics Source Code Library, record ascl:1102.026, Feb., 2011.

\bibitem{2020SciPy-NMeth}
P.~Virtanen, R.~Gommers, T.E.~Oliphant, M.~Haberland, T.~Reddy and D.e.a.~Cournapeau, \emph{{{SciPy} 1.0: Fundamental Algorithms for Scientific Computing in Python}}, \href{https://doi.org/10.1038/s41592-019-0686-2}{\emph{Nature Methods} {\bfseries 17} (2020) 261}.

\bibitem{10.7717/peerj-cs.103}
A.~Meurer, C.P.~Smith and M.e.a.~Paprocki, \emph{Sympy: symbolic computing in python}, \href{https://doi.org/10.7717/peerj-cs.103}{\emph{PeerJ Computer Science} {\bfseries 3} (2017) e103}.

\bibitem{Hunter:2007}
J.D.~Hunter, \emph{Matplotlib: A 2d graphics environment}, \href{https://doi.org/10.1109/MCSE.2007.55}{\emph{Computing in Science \& Engineering} {\bfseries 9} (2007) 90}.

\bibitem{2005ApJ...622..759G}
K.M.~{G{\'o}rski}, E.~{Hivon}, A.J.~{Banday}, B.D.~{Wandelt}, F.K.~{Hansen}, M.~{Reinecke} et~al., \emph{{HEALPix: A Framework for High-Resolution Discretization and Fast Analysis of Data Distributed on the Sphere}}, \href{https://doi.org/10.1086/427976}{\emph{\apj} {\bfseries 622} (2005) 759} [\href{https://arxiv.org/abs/arXiv:astro-ph/0409513}{{\ttfamily arXiv:astro-ph/0409513}}].

\end{thebibliography}\endgroup








\end{document}